\documentclass[twocolumn,aps,pra,floatfix]{revtex4}
\usepackage{graphicx}
\usepackage{times}
\usepackage{nicefrac}
\usepackage{amsmath}
\usepackage{amsfonts}
\usepackage{amssymb}
\usepackage{amsthm}
\usepackage{epsf}
\usepackage{bm}
\usepackage{bbm}
\usepackage{times}

\usepackage{dcolumn}

\newcolumntype{.}{D{x}{}{-1}}
\newcolumntype{w}[1]{D{.}{.}{#1}}

\begin{document}

\newcommand{\half}{\frac12}
\newcommand{\vare}{\varepsilon}
\newcommand{\eps}{\epsilon}
\newcommand{\pr}{^{\prime}}
\newcommand{\ppr}{^{\prime\prime}}
\newcommand{\pp}{{p^{\prime}}}
\newcommand{\ppp}{{p^{\prime\prime}}}
\newcommand{\hp}{\hat{\bfp}}
\newcommand{\hr}{\hat{\bfr}}
\newcommand{\hk}{\hat{\bfk}}
\newcommand{\hx}{\hat{\bfx}}
\newcommand{\hpp}{\hat{\bfpp}}
\newcommand{\hq}{\hat{\bfq}}
\newcommand{\rqq}{{\rm q}}
\newcommand{\bfk}{{\bm{k}}}
\newcommand{\bfp}{{\bm{p}}}
\newcommand{\bfq}{{\bm{q}}}
\newcommand{\bfr}{{\bm{r}}}
\newcommand{\bfx}{{\bm{x}}}
\newcommand{\bfy}{{\bm{y}}}
\newcommand{\bfz}{{\bm{z}}}
\newcommand{\bfpp}{{\bm{\pp}}}
\newcommand{\bfppp}{{\bm{\ppp}}}
\newcommand{\balpha}{\bm{\alpha}}
\newcommand{\bvare}{\bm{\vare}}
\newcommand{\bgamma}{\bm{\gamma}}
\newcommand{\bGamma}{\bm{\Gamma}}
\newcommand{\bLambda}{\bm{\Lambda}}
\newcommand{\bmu}{\bm{\mu}}
\newcommand{\bnabla}{\bm{\nabla}}
\newcommand{\bvarrho}{\bm{\varrho}}
\newcommand{\bsigma}{\bm{\sigma}}
\newcommand{\bTheta}{\bm{\Theta}}
\newcommand{\bphi}{\bm{\phi}}
\newcommand{\bomega}{\bm{\omega}}
\newcommand{\intzo}{\int_0^1}
\newcommand{\intinf}{\int^{\infty}_{-\infty}}
\newcommand{\lbr}{\langle}
\newcommand{\rbr}{\rangle}
\newcommand{\ThreeJ}[6]{
        \left(
        \begin{array}{ccc}
        #1  & #2  & #3 \\
        #4  & #5  & #6 \\
        \end{array}
        \right)
        }
\newcommand{\SixJ}[6]{
        \left\{
        \begin{array}{ccc}
        #1  & #2  & #3 \\
        #4  & #5  & #6 \\
        \end{array}
        \right\}
        }
\newcommand{\NineJ}[9]{
        \left\{
        \begin{array}{ccc}
        #1  & #2  & #3 \\
        #4  & #5  & #6 \\
        #7  & #8  & #9 \\
        \end{array}
        \right\}
        }
\newcommand{\Vector}[2]{
        \left(
        \begin{array}{c}
        #1     \\
        #2     \\
        \end{array}
        \right)
        }

\newcommand{\Dmatrix}[4]{
        \left(
        \begin{array}{cc}
        #1  & #2   \\
        #3  & #4   \\
        \end{array}
        \right)
        }
\newcommand{\Dcase}[4]{
        \left\{
        \begin{array}{cl}
        #1  & #2   \\
        #3  & #4   \\
        \end{array}
        \right.
        }
\newcommand{\cross}[1]{#1\!\!\!/}

\newcommand{\Za}{{Z \alpha}}
\newcommand{\im}{{ i}}

\title{Two-photon exchange corrections to the $\bm{g}$ factor of Li-like ions}

\author{V. A. Yerokhin}
\affiliation{Center for Advanced Studies,
        Peter the Great St.~Petersburg Polytechnic University, Polytekhnicheskaya 29,
        St.~Petersburg 195251, Russia}
\affiliation{Max~Planck~Institute for Nuclear Physics, Saupfercheckweg~1, D-69117 Heidelberg,
Germany}

\author{C. H. Keitel}
\affiliation{Max~Planck~Institute for Nuclear Physics, Saupfercheckweg~1, D-69117 Heidelberg,
Germany}

\author{Z. Harman}
\affiliation{Max~Planck~Institute for Nuclear Physics, Saupfercheckweg~1, D-69117 Heidelberg,
Germany}

\begin{abstract}

We report calculations of QED corrections to the $g$ factor of Li-like ions induced by the
exchange of two virtual photons between the electrons. The calculations are performed within
QED theory to all orders in the nuclear binding strength parameter $\Za$, where $Z$ is the
nuclear charge number and $\alpha$ is the fine-structure constant. In the region of low nuclear
charges we compare results from three different methods: QED, relativistic many-body
perturbation theory, and nonrelativistic QED. All three methods are shown to yield consistent
results. With our calculations we improve the accuracy of the theoretical predictions of the
$g$ factor of the ground state of Li-like carbon and oxygen by about an order of magnitude. Our
theoretical results agree with those from previous calculations but differ by 3-4 standard
deviations from the experimental results available for silicon and calcium.

\end{abstract}


\maketitle

\section{Introduction}

Modern Penning-trap experiments based on the continuous Stern-Gerlach effect provide very precise
measurements of the Zeeman splitting of energy levels in one- and few-electron ions
\cite{sturm:11,koehler:16,arapoglou:19}. The linear Zeeman splitting is usually parameterized in
terms of the $g$ factor of the atomic system. The fractional accuracy of the recent measurements
of the $g$ factors of H-like and Li-like ions has reached few parts in $10^{-11}$
\cite{sturm:14,glazov:19}. Combined with dedicated theoretical calculations, these measurements
provided the determination of the electron mass \cite{mohr:16:codata} and one of the best tests
of the bound-state quantum electrodynamics (QED) \cite{sturm:13:Si}. Extension of these tests
towards heavier ions are anticipated in the future \cite{vogel:19}, which might open new ways for
determination of the fine-structure constant $\alpha$ \cite{shabaev:06:prl,yerokhin:16:gfact:prl}
and searches for physics beyond the Standard Model \cite{debierre:20}.

In view of the very high accuracy of the measurements, theoretical investigations of atomic $g$
factors often need to be carried out without any expansion in the nuclear binding strength
parameter $\Za$ (where $Z$ is the nuclear charge number). In such calculations, the
electron-electron interaction has to be treated by perturbation theory. The starting point of the
perturbation expansion is the hydrogenic approximation, i.e., the approximation of
non-interacting electrons. The electron-correlation corrections come from the exchange of virtual
photons between the electrons. An exchange by each photon leads to the suppression of the
corresponding correction by a parameter of $1/Z$. The first-order perturbation correction
$\sim\!1/Z^1$ is due to the one-photon exchange. This correction is relatively simple and was
calculated for Li-like ions first in Ref.~\cite{shabaev:02:li} and later reproduced in
Refs.~\cite{yerokhin:16:gfact:pra,cakir:20}.

The QED calculation of the two-photon exchange correction $\sim\!1/Z^2$ is a difficult task.
First calculations of this correction were accomplished in Refs.~\cite{volotka:12,volotka:14}. In
these studies, results were reported for just four ions and their numerical uncertainty was
significant on the level of the current experimental precision. In the present work we will
perform an independent calculation of the two-photon exchange correction for the ground state of
Li-like ions. Our goals will be to cross-check the previous calculations, to improve the
numerical accuracy, and to study the $Z$-dependence of the two-photon correction in the low-$Z$
region, checking the consistency of the applied method with the $Z\alpha$-expansion calculations
performed recently in Ref.~\cite{yerokhin:17:gfact}.

The relativistic units ($\hbar=c=m=1$) and the Heaviside charge units ($ \alpha = e^2/4\pi$,
$e<0$) will be used throughout this paper.

\section{Electronic structure corrections to the $\boldsymbol{g}$ factor}

In the present work we assume the nucleus to be spinless and the electron configuration to be a
valence electron $v$ beyond a closed shell of core electrons denoted by $c$. Contributions to the
$g$ factor can be formally obtained as corrections induced by the effective magnetic interaction
\cite{yerokhin:10:sehfs}
\begin{align}
V_g(\bfr) = \frac{1}{\mu_v}\, (\bfr \times\balpha)_z\,,
\end{align}
where $\balpha$ is the vector of Dirac matrices and $\mu_v$ is the angular momentum projection of
the valence electron.

To the zeroth order in the electron-electron interaction, the $g$ factor of the ground state of a
Li-like ion is given by the expectation value of the magnetic potential $V_g$ on the hydrogenic
Dirac wave function of the valence $2s$ state. For the point nucleus, the result is known in the
closed form
\begin{align}
g_{\rm Dirac} = \lbr 2s | V_g| 2s \rbr = \frac23 \bigg( \Big[2\sqrt{1-(\Za)^2}+2\Big]^{1/2}+1\bigg)\,.
\end{align}

Corrections to the $g$ factor of a Li-like ion due to the presence of core electrons are
evaluated by perturbation theory in the electron-electron interaction, with the expansion
parameter $1/Z$. The leading correction of order $1/Z^1$ is induced by the one-photon exchange
between the valence and core electrons. The corresponding correction was calculated in
Ref.~\cite{shabaev:02:li} (see also Ref.~\cite{yerokhin:16:gfact:pra}) and can be written as
\begin{align}\label{eq0:4}
\Delta g_{\rm 1ph} = &\ 2 \sum_{\mu_{c}} \Big[  \Lambda_{\rm 1ph}(vcvc)+\Lambda_{\rm 1ph}(cvcv)
 \nonumber \\ &
                                                  -\Lambda_{\rm 1ph}(cvvc)-\Lambda_{\rm 1ph}(vccv) \Big]
\,,
\end{align}
where the summation runs over the angular-momentum projection of the core electron $\mu_c$ and
\begin{align}
\Lambda_{\rm 1ph}(abcd) = &\ \sum_{n\ne a} \frac{\lbr a | V_g | n\rbr \lbr nb| I(\Delta_{db}) | cd\rbr}{\vare_a-\vare_n}
\nonumber \\ &
+ \frac14\, \lbr ab| I'(\Delta_{db})|cd\rbr\,\Big( \lbr d|V_g|d\rbr - \lbr b|V_g|b\rbr\Big)
\,.
\end{align}
Here, $\Delta_{ab} = \vare_a-\vare_b$, $I(\omega)$ is the operator of the electron-electron
interaction, and $I'(\omega) =
\partial I(\omega)/(\partial \omega)$.

The electron-electron interaction operator $I(\omega)$ is defined as
\begin{equation}\label{a1}
  I(\omega,\bfr_{1},\bfr_{2}) = e^2\, \alpha_{1}^{\mu} \alpha_{2}^{\nu}\, D_{\mu\nu}(\omega,\bfr_{1},\bfr_2)\,,
\end{equation}
where $\alpha^{\mu}_{a} = (1,\balpha_{a})$ is the four-vector of Dirac matrices acting on
$\bfr_a$, $D_{\mu\nu}$ is the photon propagator, and $\omega$ is the photon energy. In the
present work we use the photon propagator in the Feynman and Coulomb gauges. In the Feynman
gauge, the electron-electron interaction takes the simplest form,
\begin{equation}
I_{\rm Feyn}(\omega) =  \alpha\,\big(1 - \balpha_1\cdot\balpha_2\big)\,
  \frac{e^{i|\omega|r_{12}}}{r_{12}}\,,
\end{equation}
where $r_{12} = |\bfr_{12}| =  |\bfr_{1} - \bfr_{2}|$, and $|\omega|$ should be understood as
$|\omega| = \sqrt{\omega^2+i\epsilon}$, where $\epsilon$ is a positive infinitesimal addition.
The electron-electron interaction operator in the Coulomb gauge reads
\begin{align}\label{eq:I:coul}
I_{\rm Coul}(\omega) = &\ \alpha\Biggl[
\frac1{r_{12}} - \balpha_1\cdot\balpha_2\, \frac{e^{i|\omega|r_{12}}}{r_{12}}
 \nonumber \\ &
 + \frac{\left( \balpha_1\cdot\bnabla_1\right)\left(\balpha_2\cdot\bnabla_2\right)}{\omega^2}
 \frac{e^{i|\omega|r_{12}}-1}{r_{12}} \Biggr]\,.
\end{align}

\begin{figure}
\centerline{
\resizebox{0.5\textwidth}{!}{%
  \includegraphics{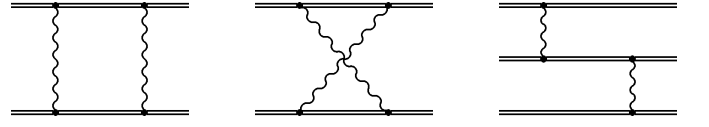}
}
}
 \caption{Feynman diagrams representing the two-photon exchange correction to the Lamb shift.
The three graphs are referred to, from left to right,  as the ladder, the crossed, and the
three-electron diagrams, respectively. The double line denotes the electron propagating in
the field of the nucleus, the wavy line denotes the virtual photon.
\label{fig:Feyn:Lamb}}
\end{figure}

\begin{figure}
\centerline{
\resizebox{0.245\textwidth}{!}{%
  \includegraphics{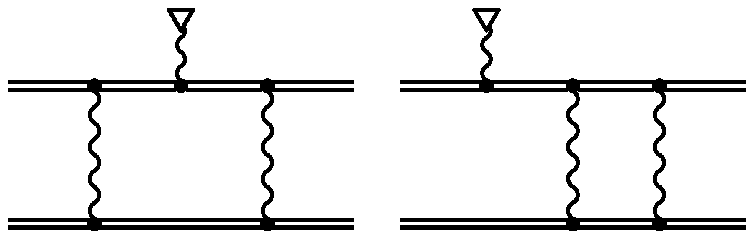}
}
\hspace*{0.05cm}
\resizebox{0.245\textwidth}{!}{%
  \includegraphics{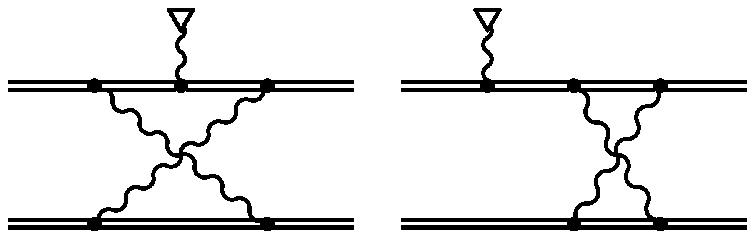}
}
}\vspace*{0.1cm}
\centerline{
\resizebox{0.5\textwidth}{!}{%
  \includegraphics{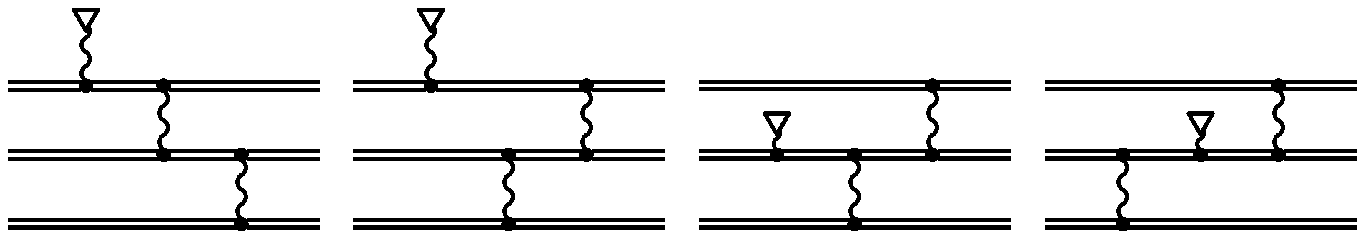}
}
}
 \caption{Feynman diagrams representing the two-photon exchange corrections to the $g$ factor. The
 wavy line terminated by a cross denotes the magnetic interaction.
\label{fig:Feyn:gfact}}
\end{figure}

It can be easily seen that the one-photon exchange correction $\Delta g_{\rm 1ph}$ given by
Eq.~(\ref{eq0:4}) can be obtained from the corresponding correction to the Lamb shift,
\begin{align}
\Delta E_{\rm 1ph}(vc) = \sum_{\mu_c} \Big[
 \lbr cv| I(0) | cv\rbr - \lbr vc| I(\Delta_{vc}) | cv\rbr\Big]\,,
\end{align}
by perturbing this expression with the magnetic potential $V_g$. Specifically, one perturbs the
one-electron wave functions,
\begin{align}\label{eq0:9}
|a\rbr \to |a\rbr + |\delta a\rbr\,, \ \ |\delta a\rbr = \sum_{n \neq a} \frac{|n\rbr \lbr n| V_g|a\rbr}{\vare_a - \vare_n}\,,
\end{align}
and energies,
\begin{align}
\vare_a \to \vare_a + \delta \vare_a\,, \ \ \delta\vare_a = \lbr a | V_g| a\rbr\,.
\end{align}
In this work, we use this approach in order to obtain formulas for the two-photon exchange
corrections to the $g$ factor. We start with the two-photon exchange correction for the Lamb
shift, graphically represented in Fig.~\ref{fig:Feyn:Lamb}. The Feynman diagrams for the $g$
factor in Fig.~\ref{fig:Feyn:gfact} are obtained from the Lamb-shift diagrams by inserting the
magnetic interaction $V_g$ in all possible ways. The corresponding formulas for the $g$ factor
are obtained by using formulas for the two-photon exchange correction for the Lamb shift derived
in Refs.~\cite{shabaev:94:ttg1,yerokhin:01:2ph} and perturbing them with the magnetic potential
$V_g$.

The two-photon exchange correction to the $g$ factor is conveniently represented as a sum of the
direct (``dir''), exchange (``ex''), and the three-electron (``3el'') contributions, obtained as
perturbations of the corresponding Lamb-shift corrections. Furthermore, each of the three
contributions is sub-divided into the irreducible (``ir'') and reducible (``red'') parts. The
reducible parts are induced by the intermediate states degenerate in energy with the energy of
the reference state of the ion. We thus represent the total two-photon exchange correction to the
$g$ factor as the sum of three irreducible and three reducible contributions,
\begin{align} \label{eq0:11}
\Delta g_{\rm 2ph} = &\ \Delta g_{\rm ir, dir} + \Delta g_{\rm ir, ex} + \Delta g_{\rm 3el, ir}
 \nonumber \\ &
+ \Delta g_{\rm red, dir} + \Delta g_{\rm red, ex} + \Delta g_{\rm 3el, red}\,.
\end{align}
We now examine each of these terms one by one.

\subsection{Direct irreducible part}

The direct irreducible contribution comes from the ladder (``lad'') and crossed (``cr'')
diagrams. For the Lamb shift, this contribution is given by Eq.~(32) of
Ref.~\cite{yerokhin:01:2ph}. Changing the variable $\omega \to -\omega$ in the ladder part and
using the property $I(\omega) = I(-\omega)$, we write the expression as
\begin{align}\label{eq:1}
\Delta E_{\rm ir, dir} &\ =
   \frac{i}{2\pi} \int_{-\infty}^{\infty} d\omega\,
\Bigg[  \sum_{n_1n_2 \ne cv,vc}\!
     \frac{F_{\rm lad, dir}(\omega, n_1n_2)}
      {(\tilde{\Delta}_{cn_1}+\omega)(\tilde{\Delta}_{vn_2}-\omega)}
 \nonumber \\ & \
   +   \sum_{n_1n_2 \ne cv}
      \frac{F_{\rm cr, dir}(\omega, n_1n_2)}
      {(\tilde{\Delta}_{cn_1}-\omega)(\tilde{\Delta}_{vn_2}-\omega)}
      \Bigg]\,,
\end{align}
where $\tilde{\Delta}_{an} \equiv \vare_a - \vare_n(1-i0)$. Furthermore,
\begin{align}
F_{\rm lad, dir}(\omega, n_1n_2) &\ =
  \sum_{\mu_c \mu_1\mu_2}
     \lbr cv| I(\omega) |n_1n_2\rbr\, \lbr n_1n_2| I(\omega)| cv\rbr
     \,, \\
F_{\rm cr, dir}(\omega, n_1n_2) &\ =
  \sum_{\mu_c \mu_1\mu_2}
   \lbr cn_2| I(\omega) |n_1v\rbr\, \lbr n_1v| I(\omega)| cn_2\rbr
   \,,
\end{align}
where $\mu_1$ and $\mu_2$ denote the angular-momentum projections of the states $n_1$ and $n_2$,
respectively. The summations over $n$'s run over the complete spectrum of the Dirac equation,
implying the sum over the corresponding relativistic angular quantum numbers $\kappa_n$ and the
principal quantum numbers of the discrete spectrum and the integration over the continuum part of
the spectrum. The terms excluded from the summation over $n_{1}$ and $n_2$ in Eq.~(\ref{eq:1})
will be accounted for by the reducible part.

Formulas for the $g$ factor are obtained by perturbing the above expressions with the magnetic
potential $V_g$. One perturbs the initial-state and intermediate-state wave functions and
energies in the denominators. Perturbations of wave functions lead to the corrections $\delta
F$'s,
\begin{align}  \label{eq:6}
\delta F_{\rm lad, dir}(\omega, n_1n_2) = &\
  2\, \sum_{\mu_c \mu_1\mu_2}
   \Big[
     \lbr cv| I(\omega) |n_1n_2\rbr\, \lbr n_1n_2| I(\omega)| \delta cv\rbr
    \nonumber \\
     &
   +\lbr cv| I(\omega) |n_1n_2\rbr\, \lbr n_1n_2| I(\omega)| c \delta v\rbr
    \nonumber \\
     &
   +\lbr cv| I(\omega) |n_1n_2\rbr\, \lbr \delta n_1n_2| I(\omega)| c v\rbr
    \nonumber \\
     &
   +\lbr cv| I(\omega) |n_1n_2\rbr\, \lbr n_1 \delta  n_2| I(\omega)| c v\rbr
   \Big]\,,
\end{align}
\begin{align}
\delta F_{\rm cr, dir}(\omega, n_1n_2) = &\
  2\,\sum_{\mu_c \mu_1\mu_2}
   \Big[
   \lbr cn_2| I(\omega) |n_1v\rbr\, \lbr n_1v| I(\omega)| \delta cn_2\rbr
    \nonumber \\
     &
  + \lbr cn_2| I(\omega) |n_1 \delta v\rbr\, \lbr n_1v| I(\omega)| cn_2\rbr
    \nonumber \\
     &
  + \lbr cn_2| I(\omega) |n_1 v\rbr\, \lbr \delta n_1v| I(\omega)| cn_2\rbr
    \nonumber \\
     &
  + \lbr c \delta n_2| I(\omega) |n_1 v\rbr\, \lbr n_1v| I(\omega)| cn_2\rbr
   \Big]\,,
\end{align}
with perturbed wave functions $|\delta a\rbr$ defined by Eq.~(\ref{eq0:9}). Perturbations of
energies in the denominators leads to corrections $\delta_1F$ and $\delta_2F$,
\begin{widetext}
\begin{align}
\delta_1 F_{\rm lad, dir}(\omega, n_1n_2) = &\
  \sum_{\mu_c \mu_1\mu_2}
\big(V_{n_1n_1}-V_{cc}\big)\,
 \lbr cv| I(\omega) |n_1n_2\rbr\, \lbr n_1n_2| I(\omega)| cv\rbr
   \,, \\
\delta_2 F_{\rm lad, dir}(\omega, n_1n_2) = &\
  \sum_{\mu_c \mu_1\mu_2}
\big(V_{n_2n_2}-V_{vv}\big)\,
 \lbr cv| I(\omega) |n_1n_2\rbr\, \lbr n_1n_2| I(\omega)| cv\rbr
   \,, \\
\delta_1 F_{\rm cr, dir}(\omega, n_1n_2) = &\
  \sum_{\mu_c \mu_1\mu_2}
\big(V_{n_1n_1}-V_{cc}\big)
  \lbr cn_2| I(\omega) |n_1v\rbr\, \lbr n_1v| I(\omega)| cn_2\rbr
   \,, \\
\delta_2 F_{\rm cr, dir}(\omega, n_1n_2) = &\
  \sum_{\mu_c \mu_1\mu_2}
\big(V_{n_2n_2}-V_{vv}\big)
  \lbr cn_2| I(\omega) |n_1v\rbr\, \lbr n_1v| I(\omega)| cn_2\rbr
   \,,
\end{align}
where $V_{ab} = \lbr a|V_g|b\rbr$. Finally, the correction to the $g$ factor is
\begin{align}\label{eq:4}
\Delta g_{\rm ir, dir} = &\
   \frac{i}{2\pi} \int_{-\infty}^{\infty} d\omega\,
   \Bigg\{
  \sum_{n_1n_2 \ne cv,vc}
   \bigg[
   \frac{\delta F_{\rm lad, dir}(\omega,n_1n_2)}
      {(\tilde{\Delta}_{cn_1}+\omega)(\tilde{\Delta}_{vn_2}-\omega)}
   + \frac{\delta_1 F_{\rm lad, dir}(\omega,n_1n_2)}
      {(\tilde{\Delta}_{cn_1}+\omega)^2(\tilde{\Delta}_{vn_2}-\omega)}
   + \frac{\delta_2 F_{\rm lad, dir}(\omega,n_1n_2)}
      {(\tilde{\Delta}_{cn_1}+\omega)(\tilde{\Delta}_{vn_2}-\omega)^2}
\bigg]
 \nonumber \\ &
+  \sum_{n_1n_2 \ne cv}
   \bigg[
   \frac{\delta F_{\rm cr, dir}(\omega,n_1n_2)}
      {(\tilde{\Delta}_{cn_1}-\omega)(\tilde{\Delta}_{vn_2}-\omega)}
   + \frac{\delta_1 F_{\rm cr, dir}(\omega,n_1n_2)}
      {(\tilde{\Delta}_{cn_1}-\omega)^2(\tilde{\Delta}_{vn_2}-\omega)}
   + \frac{\delta_2 F_{\rm cr, dir}(\omega,n_1n_2)}
      {(\tilde{\Delta}_{cn_1}-\omega)(\tilde{\Delta}_{vn_2}-\omega)^2}
\bigg]
\Bigg\}\,.
\end{align}

\subsection{Exchange irreducible part}

The exchange irreducible contribution for the Lamb shift is given by (see Eq.~(32) of
Ref.~\cite{yerokhin:01:2ph})
\begin{align}\label{eq:11}
\Delta E_{\rm ir, ex} = &\
   \frac{i}{2\pi} \int_{-\infty}^{\infty} d\omega\,
\bigg[  \sum_{n_1n_2 \ne cv,vc}
     \frac{F_{\rm lad, ex}(\omega,n_1n_2)}
      {(\tilde{\Delta}_{vn_1}-\omega)(\tilde{\Delta}_{cn_2}+\omega)}
      +
        \sum_{n_1n_2 \ne cc,vv}
      \frac{F_{\rm cr, ex}(\omega,n_1n_2)}
      {(\tilde{\Delta}_{vn_1}-\omega)(\tilde{\Delta}_{vn_2}-\omega)}
\bigg]
      \,,
\end{align}
where the functions $F$ are given by
\begin{align}
F_{\rm lad, ex}(\omega,n_1n_2) &\ =
  (-1)\sum_{\mu_c \mu_{1}\mu_{2}}
     \lbr vc| I(\omega) |n_1n_2\rbr\, \lbr n_1n_2| I(\widetilde{\omega})| cv\rbr
      \,,
\\
F_{\rm cr, ex}(\omega,n_1n_2) &\ =
  (-1)\sum_{\mu_c \mu_{1}\mu_{2}}
     \lbr vn_2| I(\omega) |n_1v\rbr\, \lbr n_1c| I(\widetilde{\omega})| cn_2\rbr
      \,,
\end{align}
with $\widetilde{\omega} \equiv \omega - \Delta$ and $\Delta \equiv \vare_v -\vare_c$. Again, the
terms excluded from the summation over $n_1$ and $n_2$ will be accounted for by the corresponding
reducible part. It should be mentioned that our present definition of the irreducible part
differs slightly from that of Ref.~\cite{yerokhin:01:2ph}. Specifically, we here do {\em not}
exclude from the summation over $n_1$ and $n_2$ the state separated by the finite nuclear size
effect from the reference state ({\em i.e.}, the $2p_{1/2}$ state for $v = 2s$ state), since it
leads to unnecessary complications in the case of the $g$ factor.

The formulas for the $g$ factor are obtained similarly to the direct contribution, by perturbing
expressions for the Lamb shift with the potential $V_g$,
\begin{align}\label{eq:14}
\Delta g_{\rm ir, ex} = &\
   \frac{i}{2\pi} \int_{-\infty}^{\infty} d\omega\,
    \Bigg\{
  \sum_{n_1n_2 \ne cv,vc}
   \bigg[
   \frac{\delta F_{\rm lad, ex}(\omega,n_1n_2)}
      {(\tilde{\Delta}_{vn_1}-\omega)(\tilde{\Delta}_{cn_2}+\omega)}
   + \frac{\delta_1 F_{\rm lad, ex}(\omega,n_1n_2)}
      {(\tilde{\Delta}_{vn_1}-\omega)^2(\tilde{\Delta}_{cn_2}+\omega)}
   + \frac{\delta_2 F_{\rm lad, ex}(\omega,n_1n_2)}
      {(\tilde{\Delta}_{vn_1}-\omega)(\tilde{\Delta}_{cn_2}+\omega)^2}
\bigg]
 \nonumber \\ &
+  \sum_{n_1n_2 \ne cc,vv}
   \bigg[
   \frac{\delta F_{\rm cr, ex}(\omega,n_1n_2)}
      {(\tilde{\Delta}_{vn_1}-\omega)(\tilde{\Delta}_{vn_2}-\omega)}
   + \frac{\delta_1 F_{\rm cr, ex}(\omega,n_1n_2)}
      {(\tilde{\Delta}_{vn_1}-\omega)^2(\tilde{\Delta}_{vn_2}-\omega)}
   + \frac{\delta_2 F_{\rm cr, ex}(\omega,n_1n_2)}
      {(\tilde{\Delta}_{vn_1}-\omega)(\tilde{\Delta}_{vn_2}-\omega)^2}
      \bigg]
      \Bigg\}\,.
\end{align}
The perturbations of the $F$ functions by the magnetic potential $V_g$ are defined as follows
\begin{align}
\delta F_{\rm lad, ex}(\omega,n_1n_2) = &\
  (-2)\sum_{\mu_c \mu_1 \mu_2}
  \Big[
  \lbr vc| I(\omega) |n_1n_2\rbr\, \lbr n_1n_2| I(\tilde{\omega})| \delta cv\rbr
+ \lbr vc| I(\omega) |n_1n_2\rbr\, \lbr n_1n_2| I(\tilde{\omega})| c \delta v\rbr
 \nonumber \\ &
+ \lbr vc| I(\omega) |n_1n_2\rbr\, \lbr \delta n_1n_2| I(\tilde{\omega})| c v\rbr
+ \lbr vc| I(\omega) |n_1n_2\rbr\, \lbr n_1 \delta n_2| I(\tilde{\omega})| c v\rbr
 \nonumber \\ &
- \frac12 \big(V_{vv}-V_{cc}\big)\,
\lbr vc| I(\omega) |n_1n_2\rbr\, \lbr n_1n_2| I'(\tilde{\omega})| cv\rbr
 \Big]\,,
 \\
\delta F_{\rm cr, ex}(\omega,n_1n_2) = &\
  (-2)\sum_{\mu_c \mu_1 \mu_2}
  \Big[
  \lbr vn_2| I(\omega) |n_1v\rbr\, \lbr n_1c| I(\tilde{\omega})| \delta cn_2\rbr
+ \lbr vn_2| I(\omega) |n_1\delta v\rbr\, \lbr n_1c| I(\tilde{\omega})| cn_2\rbr
 \nonumber \\ &
+ \lbr vn_2| I(\omega) |n_1v\rbr\, \lbr \delta n_1c| I(\tilde{\omega})| cn_2\rbr
+ \lbr v\delta n_2| I(\omega) |n_1v\rbr\, \lbr n_1c| I(\tilde{\omega})| cn_2\rbr
 \nonumber \\ &
- \frac12 \big(V_{vv}-V_{cc}\big)\,
\lbr vn_2| I(\omega) |n_1v\rbr\, \lbr n_1c| I'(\tilde{\omega})| cn_2\rbr
 \Big]\,,
\end{align}
and
\begin{align}
\delta_1 F_{\rm lad, ex}(\omega,n_1n_2) = &\
  \sum_{\mu_c \mu_1 \mu_2}
\big(V_{vv} -V_{n_1n_1}\big)\,
   \lbr vc| I(\omega) |n_1n_2\rbr\, \lbr n_1n_2| I(\tilde{\omega})| cv\rbr\,,
   \\
\delta_2 F_{\rm lad, ex}(\omega,n_1n_2) = &\
  \sum_{\mu_c \mu_1 \mu_2}
\big(V_{cc} - V_{n_2n_2}\big)\,
   \lbr vc| I(\omega) |n_1n_2\rbr\, \lbr n_1n_2| I(\tilde{\omega})| cv\rbr\,,
   \\
\delta_1 F_{\rm cr, ex}(\omega,n_1n_2) = &\
  \sum_{\mu_c \mu_1 \mu_2}
\big(V_{vv} - V_{n_1n_1}\big)\,
 \lbr vn_2| I(\omega) |n_1v\rbr\, \lbr n_1c| I(\tilde{\omega})| cn_2\rbr \,,
   \\
\delta_2 F_{\rm cr, ex}(\omega,n_1n_2) = &\
  \sum_{\mu_c \mu_1 \mu_2}
\big(V_{vv} - V_{n_2n_2}\big)\,
 \lbr vn_2| I(\omega) |n_1v\rbr\, \lbr n_1c| I(\tilde{\omega})| cn_2\rbr\,.
\end{align}

\subsection{Direct reducible part}

The reducible part of the two-electron diagrams for the Lamb shift is given by Eq.~(41) of
Ref.~\cite{yerokhin:01:2ph}. Separating the direct contribution, we write it as
\begin{align}\label{eq2:31}
 \Delta E_{\rm red, dir} &\ =
\frac{-i}{4\pi}\int_{-\infty}^{\infty}
  d\omega\,
  \bigg\{
    \bigg[ \frac{1}{(\omega+i0)^2}+\frac{1}{(\omega-i0)^2}\bigg]\,F_{\rm lad, dir}(\omega,cv)
    \nonumber \\ &
   +
    \bigg[ \frac{1}{(\omega+\Delta+i0)^2}+\frac{1}{(\omega+\Delta-i0)^2}\bigg]\,
     F_{\rm lad, dir}(\omega,vc)\,
   -
    \frac{2}{(\omega-i0)^2}\, F_{\rm cr, dir}(\omega,cv)\,
    \bigg\}\,.
\end{align}
\end{widetext}
We note that the crossed $(cv)$ term in the above expression exactly coincides with the one
excluded from the summation in Eq.~(\ref{eq:1}). The ladder $(cv)$ and $(vc)$ terms in the above
expression are very similar to those excluded from the summation in Eq.~(\ref{eq:1}) but differ
by signs of $i0$. Specifically, the terms excluded from the summation over $n_{1,2}$ in
Eq.~(\ref{eq:1}) contained poles both at $\omega = i0$ and $\omega = -i0$ (or at $\omega =
-\Delta + i0$ and $\omega = -\Delta - i0$), thus ``squeezing'' the integration contour between
the two poles, causing singularities. By contrast, the ladder terms in Eq.~(\ref{eq2:31}) have
double poles, from one side of the integration contour. Therefore, the integration contour can be
``moved away'' from the pole (assuming a finite photon mass in the case of $\omega = 0$), so
there is no real singularities in Eq.~(\ref{eq2:31}).

Taking into account that the ladder $(cv)$ term and the crossed $(cv)$ term cancel each other (as
proven in Ref.~\cite{shabaev:94:ttg1}), the expression in simplified further to yield
\begin{align}\label{eq2:33}
\Delta E_{\rm red, dir} =
 -\frac{i}{2\pi}{\cal P}\intinf  d\omega\,
   \frac{1}{\omega+\Delta}\, F_{\rm lad, dir}^{\prime}(\omega,vc)
   \,,
\end{align}
where ${\cal P}$ denotes the principal value of the integral and $F^{\prime}(\omega) =
\partial F(\omega)/(\partial\omega)$.

Corrections to the $g$ factor arise through perturbations of the Lamb-shift formulas with the
magnetic potential $V_g$. We divide them into two parts,
\begin{align}
\Delta g_{\rm red, dir} = \Delta g_{\rm red, dir, wf} + \Delta g_{\rm red, dir, en}\,,
\end{align}
where the first term is induced by perturbations of the wave functions and the second, by
perturbations of the energies. The perturbed-wave-function part is immediately obtained from
Eq.~(\ref{eq2:33}) as
\begin{align}\label{eq2:35}
\Delta g_{\rm red, dir, wf} =&\   -\frac{i}{2\pi} {\cal P} \intinf   d\omega\,
   \frac1{\omega+\Delta}\, \delta F_{\rm lad, dir}^{\prime}(\omega,vc)\,,
\end{align}
where $\delta F_{\rm lad, dir}$ is defined by Eq.~(\ref{eq:6}). The derivation of the
energy-perturbed reducible part is more difficult and is carried out by perturbing the formulas
given by Eq.~(47) of Ref.~\cite{shabaev:94:ttg1}. This derivation is potentially problematic,
because vanishing contributions to the Lamb shift may induce nonzero magnetic perturbations. For
example, the energy difference $\Delta_{an} = \vare_a - \vare_n$ induces a perturbation $\lbr
a|V_g|a\rbr - \lbr n|V_g|n\rbr$. If $\vare_n = \vare_a$ and $\mu_n \neq \mu_a$, the energy
difference vanishes but the magnetic perturbation survives. In order to avoid potential
ambiguities, we fix the reducible part by requirement of the gauge invariance of the total
correction to the $g$ factor. The result for the direct energy-perturbation reducible part is
\begin{align}\label{eq2:36}
\Delta g_{\rm red, dir, en} &\  =  \frac14 \Big[
            -\delta_1 F_{\rm lad, dir}^{\prime\prime}(0,cv)
             -\delta_2 F_{\rm lad, dir}^{\prime\prime}(0,cv)
  \nonumber \\ &\
              -\delta_1 F_{\rm lad, dir}^{\prime\prime}(\Delta,vc)
             -\delta_2 F_{\rm lad, dir}^{\prime\prime}(\Delta,vc)
  \nonumber \\ &\
             + \delta_1 F^{\prime\prime}_{\rm cr, dir}(0,cv)
             +\delta_2 F^{\prime\prime}_{\rm cr, dir}(0,cv)
\Big]
    \nonumber \\ &\
+ \frac{i}{4\pi} {\cal P} \intinf
  d\omega\,
    \frac{\delta_1 F_{\rm lad, dir}^{\prime\prime}(\omega,vc)
          -\delta_2 F_{\rm lad, dir}^{\prime\prime}(\omega,vc)}{\omega+\Delta}
\,,
\end{align}
where $F^{\prime\prime}(\omega) =\partial^2 F(\omega)/(\partial\omega)^2$.

It should be pointed out that the second integration by parts, leading to the second derivative
of the photon exchange operator $I''(\omega)$ in Eq.~(\ref{eq2:36}) is potentially troublesome.
The reason is that the imaginary part of the first derivative $I'(\omega)$ is discontinuous at
$\omega  = 0$. Specifically, ${\rm Im}\big[I'(0_+)\big] = -{\rm Im}\big[I'(0_-)\big] \ne 0$. This
discontinuity leads, in principle, to appearance of additional off-integral terms in
Eq.~(\ref{eq2:36}). We found, however, that their numerical contributions are completely
negligible for the case under consideration in the present paper. The same holds for the exchange
reducible part.

\begin{widetext}
\subsection{Exchange reducible part}

The reducible exchange correction for the Lamb shift is given by Eq.~(41) of
Ref.~\cite{yerokhin:01:2ph}. We write this correction as
\begin{align}
\Delta E_{\rm red, ex} = \frac{i}{2\pi}\int_{-\infty}^{\infty}
  d\omega\,
  \Bigg\{   & \
   -\frac12\, F_{\rm lad, ex}(\omega,cv)\,
    \bigg[ \frac{1}{(\Delta-\omega-i0)^2}+\frac{1}{(\Delta-\omega+i0)^2}\bigg]
    \nonumber \\ &
   -\frac12\, F_{\rm lad, ex}(\omega,vc)\,
    \bigg[ \frac{1}{(-\omega-i0)^2}+\frac{1}{(-\omega+i0)^2}\bigg]
    \nonumber \\ &
   +\, F_{\rm cr, ex}(\omega,cc)\,
    \frac{1}{(\Delta-\omega+i0)^2}
   +\, F_{\rm cr, ex}(\omega,vv)\,
    \frac{1}{(-\omega+i0)^2}
    \Bigg\}\,.
\end{align}
We observe that the crossed $(cc)$ and $(vv)$ terms in the above expression exactly coincide with
the two terms excluded from the summation in Eq.~(\ref{eq:11}). The ladder $(cv)$ and $(vc)$
terms in the above expression are similar to those excluded from the summation in
Eq.~(\ref{eq:11}) but differ from them by the signs of $i0$. We evaluate the above expression by
integrating by parts and taking the principal value of the integral, separating the pole
contribution. The result is
\begin{align}
\Delta E_{\rm red, ex} = &\ -\frac12 \Big[F^{\prime}_{\rm cr, ex}(\Delta,cc)
+ F^{\prime}_{\rm cr, ex}(0,vv)\Big]
    \nonumber \\ &
+ \frac{i}{2\pi} {\cal P} \intinf
  d\omega\,
  \Bigg[
    \frac{F_{\rm lad, ex}^{\prime}(\omega,cv)}{\Delta-\omega}\,
   + \frac{F_{\rm lad, ex}^{\prime}(\omega,vc)}{-\omega}\,
   - \frac{F_{\rm cr, ex}^{\prime}(\omega,cc)}{\Delta-\omega}\,
   - \frac{F_{\rm cr, ex}^{\prime}(\omega,vv)}{-\omega}\,
    \Bigg]\,.
\end{align}
It should be mentioned that the individual terms in the brackets under the integral in the above
formula contain singularities at $\omega = 0$ and $\omega = \Delta$. When the ladder and exchange
terms are combined together, however, the singularities disappear and the principal value of the
resulting integral becomes well defined and can be calculated numerically.

The reducible exchange contribution for the $g$ factor is the sum of perturbations of the wave
functions and  perturbations of the energies, $\Delta g_{\rm red, ex} = \Delta g_{\rm red, ex, wf
}+ \Delta g_{\rm red, ex, en}$, where
\begin{align}
\Delta g_{\rm red, ex, wf} = &\ -\frac12 \Big[\delta F^{\prime}_{\rm cr, ex}(\Delta,cc)
+ \delta F^{\prime}_{\rm cr, ex}(0,vv)\Big]
    \nonumber \\ &
+ \frac{i}{2\pi} {\cal P} \intinf
  d\omega\,
  \Bigg[
    \frac{\delta F_{\rm lad, ex}^{\prime}(\omega,cv)}{\Delta-\omega}\,
   + \frac{\delta F_{\rm lad, ex}^{\prime}(\omega,vc)}{-\omega}\,
   - \frac{\delta F_{\rm cr, ex}^{\prime}(\omega,cc)}{\Delta-\omega}\,
   - \frac{\delta F_{\rm cr, ex}^{\prime}(\omega,vv)}{-\omega}\,
    \Bigg]\,,
\end{align}
\begin{align}
\Delta g_{\rm red, ex, en} = &\
\frac14 \bigg[-\delta_1 F_{\rm lad, ex}^{\prime\prime}(\omega,cv)
              -\delta_2 F_{\rm lad, ex}^{\prime\prime}(\omega,cv)
              -\delta_1 F_{\rm lad, ex}^{\prime\prime}(\omega,vc)
              -\delta_2 F_{\rm lad, ex}^{\prime\prime}(\omega,vc)
    \nonumber \\ &
    \ \ \ \ \ \
            +\delta_1 F^{\prime\prime}_{\rm cr, ex}(\Delta,cc)
            +\delta_2 F^{\prime\prime}_{\rm cr, ex}(\Delta,cc)
            +\delta_1 F^{\prime\prime}_{\rm cr, ex}(0,vv)
            +\delta_2 F^{\prime\prime}_{\rm cr, ex}(0,vv)\bigg]
    \nonumber \\ &
+ \frac{i}{4\pi} {\cal P} \intinf
  d\omega\,
  \Bigg[
    \frac{\delta_1 F_{\rm lad, ex}^{\prime\prime}(\omega,cv)
          -\delta_2 F_{\rm lad, ex}^{\prime\prime}(\omega,cv)}{\omega-\Delta}
   + \frac{\delta_1 F_{\rm lad, ex}^{\prime\prime}(\omega,vc)
          -\delta_2 F_{\rm lad, ex}^{\prime\prime}(\omega,vc)}{\omega}
    \nonumber \\ &
    \ \ \ \ \ \ \ \ \ \ \ \ \ \ \ \ \  \ \ \ \ \ \ \ \ \ \  \ \ \ \ \ \ \
   - \frac{ \delta_1 F_{\rm cr, ex}^{\prime\prime}(\omega,cc)
          +\delta_2 F_{\rm cr, ex}^{\prime\prime}(\omega,cc)}{\omega-\Delta}
   - \frac{ \delta_1 F_{\rm cr, ex}^{\prime\prime}(\omega,vv)
          +\delta_2 F_{\rm cr, ex}^{\prime\prime}(\omega,vv)}{\omega}
    \Bigg]\,.
\end{align}

\subsection{Three-electron part}

The three-electron irreducible contribution to the Lamb shift is given by Eq. (14) of
Ref.~\cite{yerokhin:01:2ph},
\begin{align} \label{eq0a}
\Delta E_{\rm 3el, ir} = \sum_{PQ} (-1)^{P+Q} {\sum_{n}}^{\prime} \
 \frac{I_{P2\,P3\,n\,Q3}(\Delta_{P3Q3})\,
       I_{P1\,n\,Q1\,Q2}(\Delta_{Q1P1})}
    {\vare_{Q1}+\vare_{Q2}-\vare_{P1}-\vare_n} \,.
\end{align}
Here, ``1'', ``2'', and ``3'' label the three electrons of the ions (in arbitrary order), the
operators $P$ and $Q$ permute the initial-state and the final-state electrons, $(-1)^P$ and
$(-1)^Q$ are the sign of permutations, and the prime on the sum means that the terms with the
vanishing denominator are excluded from the summation. Furthermore, $\Delta_{ab} \equiv \vare_a -
\vare_b$, and $I_{abcd}(\Delta) \equiv \lbr ab|I(\Delta)|cd\rbr$.

The corrections to the $g$ factor are obtained as first-order perturbations of Eq.~(\ref{eq0a})
by $V_g$. It is convenient to split the whole contribution into the perturbations of the external
wave functions (``pwf''), external energies (``en''), and the propagator (``ver''),
\begin{align}\label{eq0b0}
\Delta g_{\rm 3el, ir} = \Delta g^{\rm 3el}_{\rm ir, pwf}  + \Delta g^{\rm 3el}_{\rm ir, en} + \Delta g^{\rm 3el}_{\rm ir, ver}\,,
\end{align}
\begin{align} \label{eq0b}
\Delta g^{\rm 3el}_{\rm ir, pwf} = &\ 2 \sum_{PQ} (-1)^{P+Q} {\sum_{n}}^{\prime} \
 \frac1{\vare_{Q1}+\vare_{Q2}-\vare_{P1}-\vare_n}
 \biggl[
  I_{P2\,P3\,n\,\delta Q3}(\Delta_{P3Q3})\,I_{P1\,n\,Q1\,Q2}(\Delta_{Q1P1})
 \nonumber \\ &
+ I_{P2P3nQ3}(\Delta_{P3Q3})\,I_{P1\,n\,\delta Q1\,Q2}(\Delta_{Q1P1})
+ I_{P2P3nQ3}(\Delta_{P3Q3})\,I_{P1\,n\,Q1\,\delta Q2}(\Delta_{Q1P1})
  \biggr]
     \ ,
\end{align}
\begin{align} \label{eq0d}
&\ \Delta g^{\rm 3el}_{\rm ir, en}   =\sum_{PQ} (-1)^{P+Q} {\sum_{n}}^{\prime} \
\biggl[
- \left( V_{Q1Q1}  + V_{Q2Q2} - V_{P1P1} \right)
 \frac{I_{P2\,P3\,n\,Q3}(\Delta_{P3Q3})\,
       I_{P1\,n\,Q1\,Q2}(\Delta_{Q1P1})} {(\vare_{Q1}+\vare_{Q2}-\vare_{P1}-\vare_n)^2}
 \nonumber \\ &
+
 \frac{\left( V_{P3P3}  - V_{Q3Q3} \right)\, I'_{P2P3nQ3}(\Delta_{P3Q3})\,
       I_{P1\,n\,Q1\,Q2}(\Delta_{Q1P1})
       +
       \left( V_{Q1Q1}  - V_{P1P1} \right)
       I_{P2\,P3\,n\,Q3}(\Delta_{P3Q3})\,
       I'_{P1\,n\,Q1\,Q2}(\Delta_{Q1P1})
       } {\vare_{Q1}+\vare_{Q2}-\vare_{P1}-\vare_n}
       \biggr]\,,
\end{align}
\begin{align} \label{eq0c}
\Delta g^{\rm 3el}_{\rm ir, ver} = \sum_{PQ} (-1)^{P+Q} {\sum_{n_1n_2}} \,\Xi\,\,
 \frac{I_{P2\,P3\,n_1\,Q3}(\Delta_{P3Q3}) \,V_{n_1n_2}\,
       I_{P1\,n_2\,Q1\,Q2}(\Delta_{Q1P1})}
    {(\vare_{Q1}+\vare_{Q2}-\vare_{P1}-\vare_{n_1})(\vare_{Q1}+\vare_{Q2}-\vare_{P1}-\vare_{n_2})} \ ,
\end{align}
where the operator $\Xi$ acts on energy denominators $\Delta_1$, $\Delta_2$ as follows:
\begin{eqnarray} \label{eqII17}
\Xi \, \frac{X}{\Delta_1\,\Delta_2} = \left\{
   \begin{array}{cl}
 \displaystyle       \frac{ X}{ \Delta_1\Delta_2}\,,
        & \mbox{if}\         \Delta_1\ne0 \ \mbox{and}\ \Delta_2\ne0\,,\\[0.35cm]
        \displaystyle -\frac{ X}{\Delta_1^2}\,, &\mbox{if}\
                             \Delta_1\ne0 \ \mbox{and}\ \Delta_2=0\,,\\[0.35cm]
        \displaystyle  -\frac{X}{\Delta_2^2}\,, &\mbox{if}\
                             \Delta_1=0 \ \mbox{and}\ \Delta_2\ne0\,,\\[0.35cm]
        0\,, &\mbox{if}\
                             \Delta_1=0 \ \mbox{and}\ \Delta_2=0\,.\\
   \end{array}
   \right.
\end{eqnarray}

The three-electron reducible correction to the Lamb shift is given by Eq.~(19) of
Ref.~\cite{yerokhin:01:2ph},
\begin{align} \label{eq2a}
\Delta {E}_{\rm 3el, red} =&\ \frac12 \sum_{PQ} (-1)^{P+Q} \sum_{{\mu_n}\atop{\vare_n = \vare_{Q1}+\vare_{Q2}-\vare_{P1}}}\!\!\!\!\!\!
    \bigg[
    I_{P2\,P3\,n\,Q3}\pr(\Delta_{P3Q3})\,I_{P1\,n\,Q1\,Q2}(\Delta_{Q1P1})
    +I_{P2\,P3\,n\,Q3}(\Delta_{P3Q3})\,I_{P1\,n\,Q1\,Q2}\pr(\Delta_{Q1P1})
    \bigg] \,.
\end{align}
The corresponding corrections to the $g$ factor arise as perturbations of the wave functions and
energies,
\begin{align} \label{eq2b}
\Delta {g}_{\rm 3el, red} =&\ \sum_{PQ} (-1)^{P+Q}  \sum_{{\mu_n}\atop{\vare_n = \vare_{Q1}+\vare_{Q2}-\vare_{P1}}}\!\!\!\!\!\!
    \biggl\{
    I_{P2\,P3\,n\,Q3}\pr(\Delta_{P3Q3})\,I_{P1\,n\,\delta Q1\,Q2}(\Delta_{Q1P1}) + I_{P2\,P3\,n\,Q3}(\Delta_{P3Q3})\,I_{P1\,n\,\delta Q1\,Q2}\pr(\Delta_{Q1P1})
\nonumber \\ &
+   I_{P2\,P3\,n\,Q3}\pr(\Delta_{P3Q3})\,I_{P1\,n\,Q1\,\delta Q2}(\Delta_{Q1P1}) + I_{P2\,P3\,n\,Q3}(\Delta_{P3Q3})\,I_{P1\,n\,Q1\,\delta Q2}\pr(\Delta_{Q1P1})
\nonumber \\ &
+   I_{P2\,P3\,n\,\delta Q3}\pr(\Delta_{P3Q3})\,I_{P1\,n\,Q1\,Q2}(\Delta_{Q1P1}) + I_{P2\,P3\,n\,\delta Q3}(\Delta_{P3Q3})\,I_{P1\,n\,Q1\,Q2}\pr(\Delta_{Q1P1})
\nonumber \\ &
+   I_{P2\,P3\,n\,Q3}\pr(\Delta_{P3Q3})\,I_{P1\,\delta n\,Q1\,Q2}(\Delta_{Q1P1}) + I_{P2\,P3\,n\,Q3}(\Delta_{P3Q3})\,I_{P1\,\delta n\,Q1\,Q2}\pr(\Delta_{Q1P1})
\nonumber \\ &
+ (V_{P3P3}-V_{Q3Q3})
\,\Bigl[ I\ppr_{P2\,P3\,n\,Q3}(\Delta_{P3Q3})I_{P1\,n\, Q1\,Q2}(\Delta_{Q1P1})
 + I\pr_{P2\,P3\,n\,Q3}(\Delta_{P3Q3})I_{P1\,n\, Q1\,Q2}\pr(\Delta_{Q1P1})\Big]
\nonumber \\ &
+
(V_{Q1Q1}-V_{P1P1}) \,
\Big[ I\pr_{P2\,P3\,n\,Q3}(\Delta_{P3Q3})I\pr_{P1\,n\, Q1\,Q2}(\Delta_{Q1P1})
 + I_{P2\,P3\,n\,Q3}(\Delta_{P3Q3})I\ppr_{P1\,n\, Q1\,Q2}(\Delta_{Q1P1})\Big]
    \biggr\}
    \, .
\end{align}

\section{MBPT approximation}

In this section we obtain formulas for the two-photon exchange correction to the $g$ factor in
the approximation of the relativistic many-body perturbation theory (MBPT). The corresponding
formulas can be obtained from the QED expressions by (i) using the Coulomb gauge in the photon
propagators and neglecting the energy dependence, $I(\omega) \to I_{\rm Coul}(0)$, and (ii)
restricting the summations over the Dirac spectrum to the positive-energy part. Under these
assumptions, all reducible contributions vanish and the integrals over $\omega$ can be performed
by the Cauchy theorem. The $\omega$ integral for the crossed contribution vanishes, so the total
two-electron correction comes only from the ladder irreducible part. Performing the $\omega$
integrations in Eqs.~(\ref{eq:4}) and (\ref{eq:14}), we obtain the two-electron contribution in
the MBPT approximation as
\begin{align}  \label{eq3a}
\Delta g_{\rm 2el}^{\rm MBPT}
  = &\  \left.\sum_{\underset{\vare_{n_1},\vare_{n_2} > 0}{n_1n_2}}\right.^{\!\!\!\!\!\!\prime}\,\,\,
 \Bigg[ \frac{\delta F_{\rm lad, dir}(0,n_1n_2)+ \delta F_{\rm lad, ex}(0,n_1n_2)}{\vare_v+\vare_c-\vare_{n_1}-\vare_{n_2}}
\nonumber \\ &
+
\frac{\delta_1 F_{\rm lad, dir}(0,n_1n_2)+ \delta_2 F_{\rm lad, dir}(0,n_1n_2)
+ \delta_1 F_{\rm lad, ex}(0,n_1n_2)+\delta_2 F_{\rm lad, ex}(0,n_1n_2)}{(\vare_v+\vare_c-\vare_{n_1}-\vare_{n_2})^2}
\Bigg]\,.
\end{align}
\end{widetext}
Here, the prime on the sum means that the terms with the vanishing denominator should be omitted,
and the summation over $n_1$ and $n_2$ is performed over the positive-energy part of the Dirac
spectrum. We note that Eq.~(\ref{eq3a}) can be also obtained directly by  perturbing the
two-photon MBPT correction for the Lamb shift, given by Eq.~(43) of Ref.~\cite{yerokhin:01:2ph}.
The three-electron MBPT correction is immediately obtained from Eqs.~(\ref{eq0b0})-(\ref{eq0c}),
after the substitution $I(\omega) \to I_{\rm Coul}(0)$ and the restriction of the summations to
the positive-energy part of the spectrum.

We note that the standard formulation of MBPT assumes the restriction of all summations over the
Dirac spectrum to the positive-energy part. The consistent treatment of the negative-energy
spectrum is possible only within the QED theory. However, it can be easily observed that one can
include some negative-energy contributions already in the MBPT formulas, namely, in those cases
when it does not lead to the so-called continuum dissolution, i.e., vanishing energy
denominators. Specifically, one can include the negative-energy spectrum in the three-electron
contributions, Eqs.~(\ref{eq0b0})-(\ref{eq0c}), and in the magnetic perturbations of the wave
functions, Eq.~(\ref{eq0:9}). We will refer to this variant of the MBPT as ``MBPT-neg''. We will
demonstrate that such partial inclusion of the negative-energy spectrum within MBPT is crucially
important to approximately reproduce the QED results in the region of small nuclear charges,
whereas the standard MBPT yields a very much different result, even in the limit of $Z\to 0$.
Previously the same conclusion was drawn by the St.~Petersburg group
\cite{volotka:priv,wagner:13}.

The connection between the QED and MBPT formulas was extensively used in this work for checking
the numerical procedure for the $\omega$ integrations. Specifically, after neglecting the energy
dependence of the photon propagators in the Coulomb gauge, we checked that the numerical $\omega$
integration yields the same result as the analytical integration by the Cauchy theorem.

\section{Numerical evaluation}

\begin{figure*}
\centerline{
\resizebox{0.45\textwidth}{!}{%
  \includegraphics{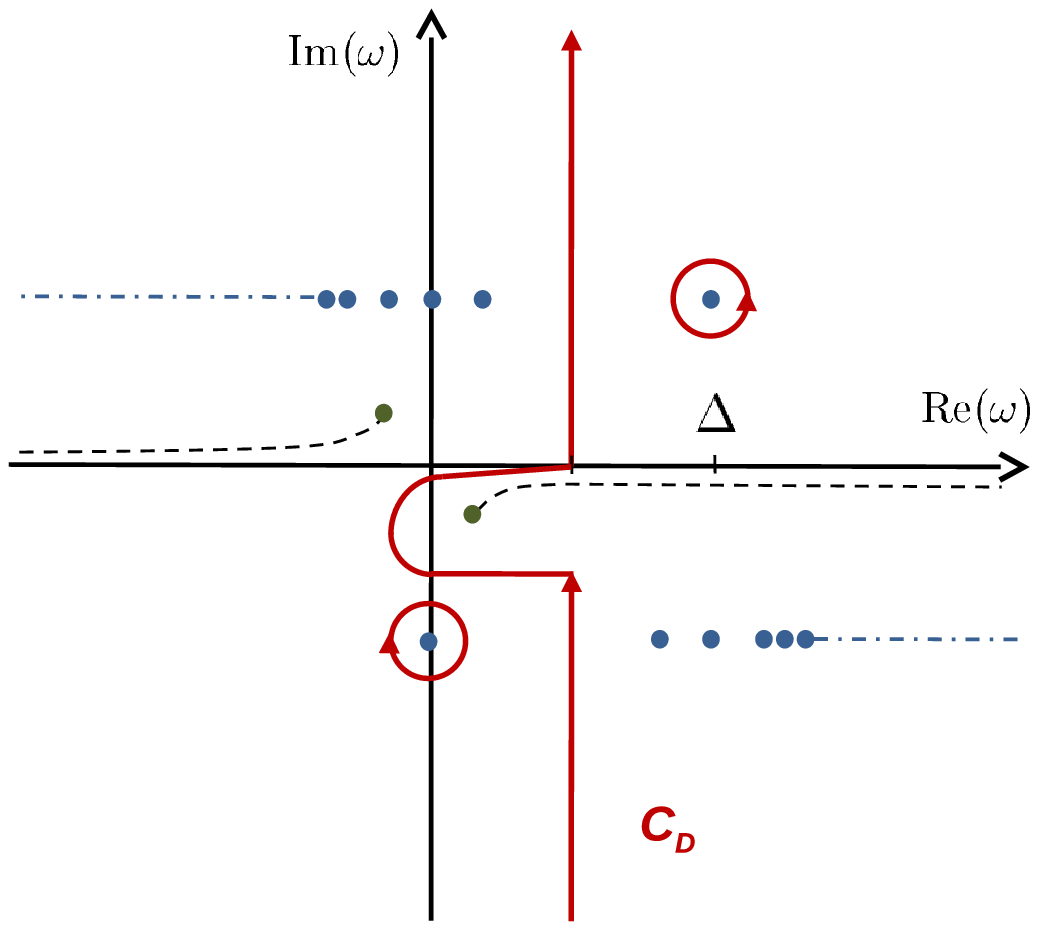}
}
\hspace*{0.5cm}
\resizebox{0.45\textwidth}{!}{%
  \includegraphics{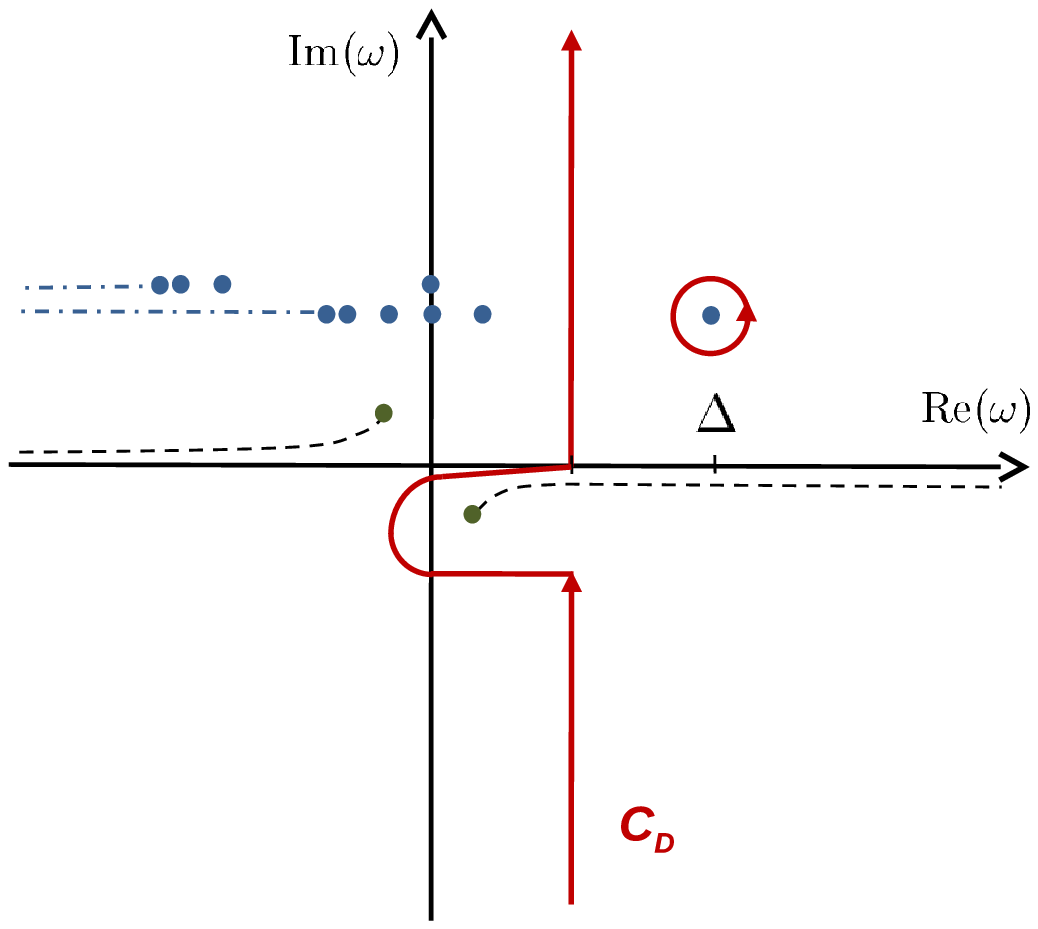}
}
}
 \caption{The singularities of the integrand in the complex $\omega$ plane and the integration contour $C_D$, for
 the ladder direct contribution (left panel) and the crossed direct contribution (right panel).
\label{fig:CD}}
\end{figure*}

\begin{figure*}
\centerline{
\resizebox{0.45\textwidth}{!}{%
  \includegraphics{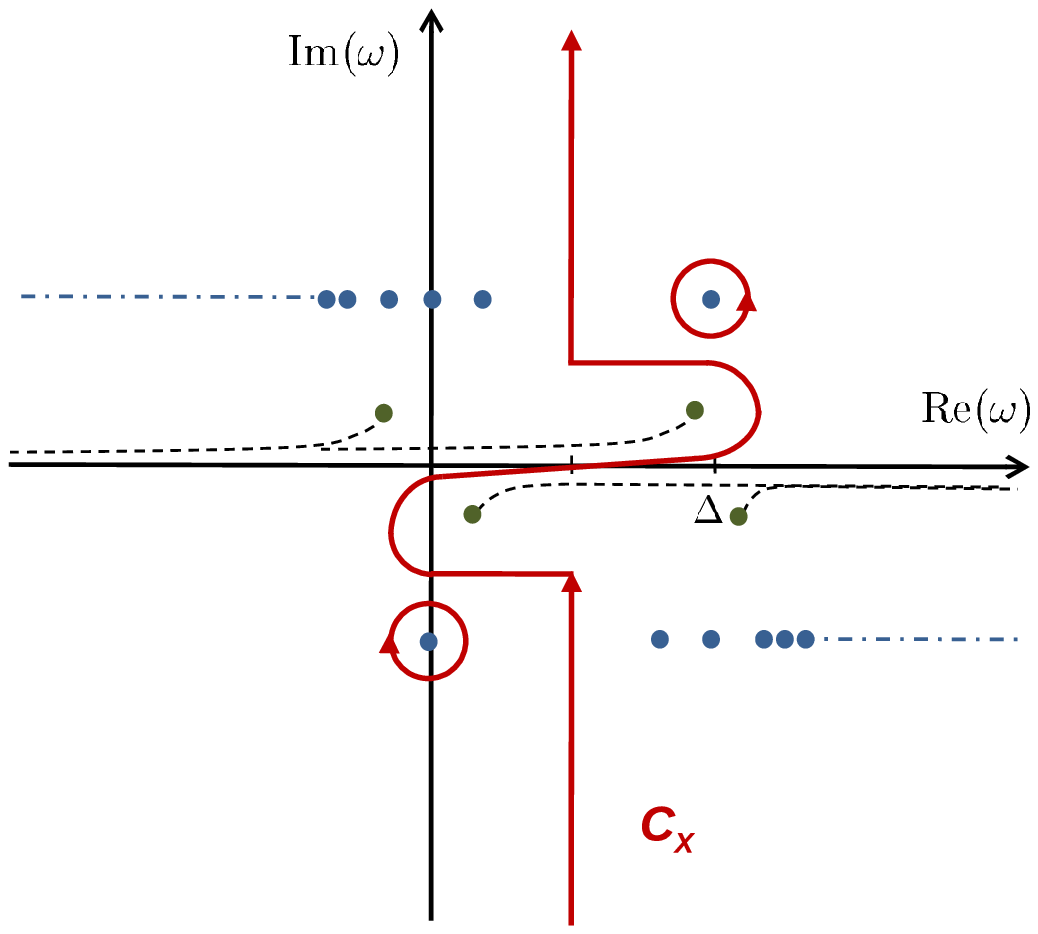}
}
\hspace*{0.5cm}
\resizebox{0.45\textwidth}{!}{%
  \includegraphics{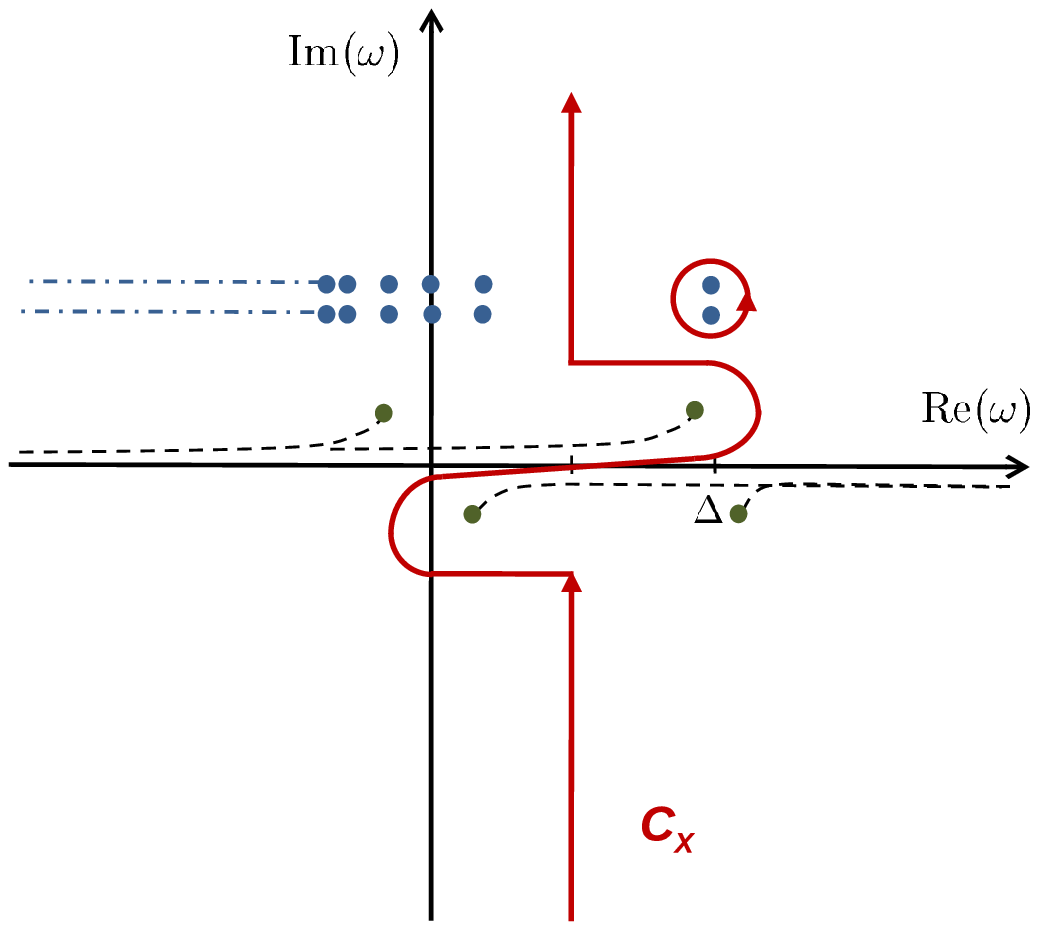}
}
}
 \caption{The contour $C_X$ and singularities of the integrand in the complex $\omega$-plane, for
 the ladder exchange contribution (left panel) and the crossed exchange contribution (right panel).
\label{fig:CX}}
\end{figure*}

We now turn to the numerical evaluation of the two-photon exchange corrections. Since the
calculation of the three-electron contributions is relatively straightforward, we concentrate
mainly on the two-electron terms. The direct and exchange irreducible contributions given by
Eqs.~(\ref{eq:4}) and (\ref{eq:14}) represent the main computational difficulty. It is
advantageous to deform the contour of the $\omega$ integration in them, in order to escape strong
oscillations of the photon propagators $\propto e^{i|\omega|r_{12}}$ for large real values of
$\omega$. Deforming the contour, one needs to take into account the branch cuts of the photon
propagators and the pole structure of the Dirac propagators. The analytical structure of the
integrand as a function of complex $\omega$ is shown in Fig.~\ref{fig:CD} for the direct part and
in Fig.~\ref{fig:CX} for the exchange part, respectively.

For the evaluation of the direct irreducible contribution, we use two different choices of the
$\omega$-integration contour. The first choice is the standard Wick rotation, $\omega \to
i\omega$, which splits the correction into the pole contribution and the integral along the
imaginary $\omega$ axis. This contour was used in the previous Lamb-shift calculations
\cite{yerokhin:01:2ph,mohr:00:pra}. The advantage of this choice is that the analysis of the pole
terms is the simplest. There are, however, also some difficulties. The first problem is the
rapidly-varying structure of the integrand in the vicinity of $\omega = 0$, due to poles of the
electron propagators lying near the imaginary axis. The second difficulty is that the
contributions with $n_2 = v$ in Eq.~(\ref{eq:4}) contain singular terms $\sim 1/\omega^2$, which
need to be integrated by parts before the numerical evaluation.

In order to achieve a more regular behaviour of the integrand for small $\omega$, we adopted the
contour $C_D$, shown in Fig.~\ref{fig:CD}. This contour is convenient for the numerical
evaluation, especially for low values of $Z$. Its disadvantage is the presence of a pole on the
low-energy part of the integration contour and thus the need to evaluate the principal value of
the integral. We find, however, that the contour $C_D$ is very similar to the contour $C_X$ used
in the evaluation of the exchange contribution and discussed in detail below. Because of this
similarity, we were able to use essentially the same numerical procedure both for the direct and
the exchange part. We checked that our numerical evaluation of the integral along the contour
$C_D$ leads to the same results as the integration along the Wick-rotated contour.

For the numerical evaluation of the exchange irreducible part, we use the contour $C_X$ depicted
in Fig.~\ref{fig:CX}. This contour was suggested in Ref.~\cite{mohr:00:pra} for the Lamb shift
and later used for the $g$ factor and hyperfine structure in
Refs.~\cite{volotka:12,kosheleva:20}. As can be seen from Fig.~\ref{fig:CX}, the deformation of
the contour from $(-\infty,\infty)$ to $C_X$ leads to appearance of pole terms at $\omega = 0$
and $\omega = \Delta$. In the case of the Lamb shift, the pole terms are identified as follows,
for the ladder contribution,
\begin{align}\label{eq:29}
   &\frac{i}{2\pi} \int_{-\infty}^{\infty} d\omega\, \frac{F(\omega)}{(\tilde{\Delta}_1-\omega)(\tilde{\Delta}_2+\omega)}
    \underset{\scriptscriptstyle \Delta_1\ne -\Delta_2}{=} \frac{F(\Delta_1)}{\Delta_1+\Delta_2}\,\delta(\Delta_1-\Delta)
 \nonumber \\ &
    + \frac{F(-\Delta_2)}{\Delta_1+\Delta_2}\,\delta(\Delta_2)
   + \frac{i}{2\pi} \int_{C_X} d\omega\, \frac{F(\omega)}{(\tilde{\Delta}_1-\omega)(\tilde{\Delta}_2+\omega)}
   \,,
\end{align}
where $\tilde{\Delta}_i$ denotes $\Delta_i$ with the infinitesimal imaginary addition according
to Eqs.~(\ref{eq:4}) and (\ref{eq:14}). For  the crossed contribution, the corresponding equation
reads
\begin{align}
   &\frac{i}{2\pi} \int_{-\infty}^{\infty} d\omega\, \frac{F(\omega)}{(\tilde{\Delta}_1-\omega)(\tilde{\Delta}_2-\omega)}
    \underset{\scriptscriptstyle \Delta_1\ne \Delta_2}{=} \ \frac{F(\Delta_1)}{\Delta_2-\Delta_1}\,\delta(\Delta_1-\Delta)
 \nonumber \\ &
    + \frac{F(\Delta_2)}{\Delta_1-\Delta_2}\,\delta(\Delta_2-\Delta)
   + \frac{i}{2\pi} \int_{C_X} d\omega\, \frac{F(\omega)}{(\tilde{\Delta}_1-\omega)(\tilde{\Delta}_2-\omega)}
   \,. \label{eq:30}
\end{align}
For the $g$ factor, formulas with squared energy denominators are required, which can be obtained
by a formal differentiation of the above formulas over $\Delta_1$ and $\Delta_2$.

For a numerical evaluation, the integral over $C_X$ is represented as a sum of three pieces,
\begin{align}
\frac{i}{2\pi}\int_{C_X}d\omega &\ \frac{I(\omega)\,I(\omega-\Delta)}{(\tilde{\Delta}_1-\omega)(\tilde{\Delta}_2\pm\omega)}
 =
\nonumber \\ &
 -\frac1{\pi} \int_0^{\delta}d\omega\,
  \frac{{\rm Im}\big[I(\omega)\big]\,I(\Delta-\omega)}{(\tilde{\Delta}_1-\omega)(\tilde{\Delta}_2\pm\omega)}
\nonumber \\ &
 -\frac1{\pi} \int_{\delta}^{\Delta} d\omega\,
  \frac{I(\omega)\,{\rm Im}\big[I(\Delta-\omega)\big]}{(\tilde{\Delta}_1-\omega)(\tilde{\Delta}_2\pm\omega)}
 \nonumber \\ &
 -\frac1{\pi} \, {\rm Re}\int_{0}^{\infty} d\omega\,
   \frac{I(\delta + i\omega)\,I(\delta + i\omega-\Delta)}{(\Delta_1-\delta - i\omega)(\Delta_2\pm (\delta + i\omega))}\,,
\end{align}
where $\delta$ is a free parameter $0 < \delta < \Delta$. A typical value of $\delta = \Delta/2$
was used. An advantage of the contour $C_X$ is that the integrand has a more regular behaviour at
the end points of the intervals, $\omega = 0$ and $\omega = \Delta$, because ${\rm
Im}\,I(\omega)\propto\omega$ as $\omega \to 0$. There are, however, singularities inside the
intervals along the real axis, and thus the infinitesimal imaginary terms $i0$ should be retained
for them. For the $g$ factor, we encounter single, double, and even triple poles on the
integration contour. Specifically, for the $v = 2s$ reference state, the singularities arise from
the intermediate states $n_{1}$ and/or $n_2 = 2p_{1/2}$, whose energy $\vare_{2p_{1/2}} <
\vare_{2s}$ is separated from the reference-state energy by the finite nuclear size effect. To
deal with these singularities, we introduce subtractions obtained by expanding the integrand in
the Taylor series in the vicinity of the poles. The subtractions remove singularities and make
the integrand a regular and smooth function suitable for the numerical integration. The
subtracted terms are then re-added, with the principal-value integrals calculated analytically.
The corresponding formulas are summarized in Appendix~\ref{sec:subtractions}.

In order to check our numerical procedure, we performed calculations also by using a different
integration contour, namely, the contour $C_{\rm irr}$ suggested in Ref.~\cite{yerokhin:01:2ph}
(shown in Fig.~5 of that work). A very good agreement of numerical results obtained with two
different contours was used as a confirmation of the internal consistency of the numerical
procedure.

For the numerical evaluation of the reducible direct and exchange contributions, we used the
$\omega$-integration contour consisting of three sections: $(-\delta -i \infty,-\delta)$, $(-
\delta,\delta)$, $(\delta,\delta + i\infty)$. The parameter $\delta$ of the contour was taken to
be $\delta > \Delta$, which allowed us to evaluate the principal value of the integrals at points
$\omega = 0$ and $\omega = \pm \Delta$.

The summations over the Dirac intermediate states were performed by using the dual kinetic
balance basis-set method \cite{shabaev:04:DKB}, with the basis set constructed with the
B-splines. The standard two-parameter Fermi model was used to represent the nuclear charge
distribution, with the nuclear radii taken from Ref.~\cite{angeli:13}. The infinite partial-wave
summation over the relativistic angular momentum quantum number $\kappa$ was performed up to
$|\kappa_{\rm max}| = 25$, with the remaining tail estimated by the polynomial fitting of the
expansion terms in $1/|\kappa|$. The largest numerical uncertainty was typically induced by
convergence in the number of basis functions. Our final values were typically obtained by
performing calculations with $N = 85$ and $N = 105$ $B$-splines and extrapolating the results to
$N\to \infty$ as $\delta g = \delta g_{N = 105} + 0.93\,(\delta g_{N = 105}-\delta g_{N = 85}$),
where the numerical coefficient was obtained by analysing the convergence pattern of our
numerical results. An example of the convergence study with respect to $N$ is presented in
Table~\ref{tab:convergence}.

In the present work we perform calculations in the Feynman and the Coulomb gauge. The expressions
for the matrix elements of the electron-electron interaction in the Coulomb gauge are summarized
in Appendix~\ref{app:coul}. We note that in the present work (unlike, e.g., in
Ref.~\cite{yerokhin:01:2ph}) we use the expression for the Coulomb-gauge matrix element
[Eq.~(\ref{eq:Rcoul})] that does not rely on commutator relations for the wave functions. This
expression is valid for the general case when the wave functions in the matrix element are not
eigenfunctions of the Dirac Hamiltonian, in particular, when they are the magnetic perturbations
of the Dirac wave functions. Another advantage of this expression is that it allows a numerical
evaluation of the Coulomb-gauge radial integrals and their derivatives for very small but
nonvanishing photon energies $\omega$. The region of small but nonzero $\omega$ is usually
numerically unstable for the expressions based on the commutator relations, especially for the
second derivative of the photon propagator $I''(\omega)$.

\begin{table}
\caption{Convergence study for the direct irreducible contribution $\Delta g_{\rm ir, dir}$ for $Z = 14$
as a function of the number of
$B$-splines in the basis set $N$, in Feynman gauge, in units of $10^{-6}$.
\label{tab:convergence}}
\begin{ruledtabular}
\begin{tabular}{lw{4.6}w{4.6}}
$N$ & \multicolumn{1}{c}{$\Delta g_{\rm ir, dir}$}
                   &  \multicolumn{1}{c}{Increment}
\\ \hline\\[-5pt]
55      & -9.442\,31 & \\
70      & -9.442\,86 & -0.000\,55 \\
85      & -9.443\,08 & -0.000\,22 \\
105     & -9.443\,20 & -0.000\,12 \\
130     & -9.443\,26 & -0.000\,06 \\
Extrap. & -9.443\,31 & -0.000\,06 \\
\end{tabular}
\end{ruledtabular}
\end{table}

\begin{figure}[t]
\centerline{
\resizebox{1.1\columnwidth}{!}{%
  \includegraphics{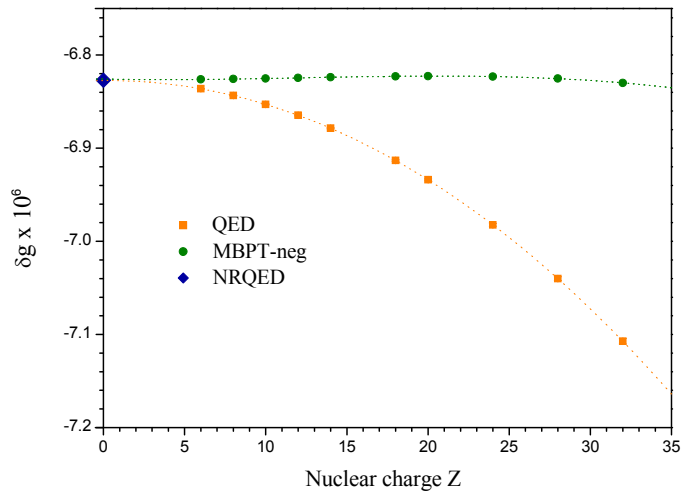}
} }
\vskip0.5cm
\caption{The two-photon exchange correction to the $g$ factor of the ground state
of Li-like ions in different approaches (QED, MBPT with the negative continuum contribution, NRQED).
The dotted line is a polynomial fit to the numerical data, to guide the eye.
\label{fig:2ph}}
\end{figure}

\begin{table*}
\caption{Individual two-photon exchange contributions to the $g$ factor of the ground state
of Li-like ions, in Feynman and Coulomb gauge. Units are $10^{-6}$.
\label{tab:2ph:Si}}
\begin{ruledtabular}
\begin{tabular}{llw{4.6}w{4.6}w{4.6}w{4.6}w{4.6}w{4.6}w{4.6}}
$Z$ & \multicolumn{1}{c}{Gauge}
                   &  \multicolumn{2}{c}{Direct}
                              &  \multicolumn{2}{c}{Exchange}
                                       &  \multicolumn{2}{c}{3-electron}
                                                  &  \multicolumn{1}{c}{Total}
                                                                   \\
                &   &  \multicolumn{1}{c}{Irred}&  \multicolumn{1}{c}{Red}
                   &  \multicolumn{1}{c}{Irred}&  \multicolumn{1}{c}{Red}
                   &  \multicolumn{1}{c}{Irred}&  \multicolumn{1}{c}{Red}\\
                                                                   \hline\\[-5pt]
14 &Feynman&    -9.4433 &   0.0015 &  -0.0739  &  0.0302  &  2.6352 &  -0.0285  &   -6.8787\,(1) \\
   &Coulomb&    -9.4417 &   0.0000 &  -0.0457  &  0.0020  &  2.6320 &  -0.0253  &   -6.8787\,(2) \\[3pt]
83 &Feynman&   -16.6698 &  -0.0053 &  -3.4213  &  1.0489  & 11.0343 &  -1.0481  &   -9.0613\,(6) \\
   &Coulomb&   -16.6175 &  -0.0631 &  -2.8158  &  0.4425  & 10.8548 &  -0.8621  &   -9.0612\,(6) \\
\end{tabular}
\end{ruledtabular}
\end{table*}

\begin{table*}
\caption{Numerical results for the two-photon exchange correction to the $g$ factor of the ground state
of Li-like ions, in units of $10^{-6}$. QED results are obtained in the Feynman gauge. ``MBPT'' labels results obtained within
the standard relativistic many-body perturbation theory. ``MBPT-neg'' labels results obtained within
MBPT supplemented by the correction from the negative-continuum part of the Dirac spectrum.
\label{tab:2ph}}
\begin{ruledtabular}
\begin{tabular}{rw{4.6}w{4.6}w{4.6}w{4.8}w{4.6}w{4.4}}
 \multicolumn{1}{c}{Z}
                   &  \multicolumn{1}{c}{Direct}
                              &  \multicolumn{1}{c}{Exchange}
                                       &  \multicolumn{1}{c}{3-electron}
                                                  &  \multicolumn{1}{c}{Total QED}
                                                  &  \multicolumn{1}{c}{MBPT-neg}
                                                  &  \multicolumn{1}{c}{MBPT}   \\
                                                                   \hline\\[-5pt]
  6   &  -9.3333  &   -0.0068  &    2.5038    &    -6.8363\,(3) & -6.8262 & -22.243 \\
  8   &  -9.3524  &   -0.0128  &    2.5216    &    -6.8436\,(2) & -6.8257 & -22.250 \\
 10   &  -9.3769  &   -0.0209  &    2.5447    &    -6.8531\,(3) & -6.8251 & -22.258 \\
 12   &  -9.4066  &   -0.0312  &    2.5730    &    -6.8648\,(2) & -6.8245 & -22.268 \\
 14   &  -9.4418  &   -0.0437  &    2.6067    &    -6.8787\,(1) & -6.8239 & -22.280 \\
      &           &            &              &    -6.876\,^a \\
 18   &  -9.5287  &   -0.0755  &    2.6909    &    -6.9133\,(2) & -6.8229 & -22.309 \\
 20   &  -9.5808  &   -0.0949  &    2.7417    &    -6.9341\,(3) & -6.8227 & -22.327 \\
 24   &  -9.7031  &   -0.1408  &    2.8615    &    -6.9824\,(3) & -6.8231 & -22.367 \\
 28   &  -9.8512  &   -0.1962  &    3.0073    &    -7.0401\,(3) & -6.8252 & -22.414 \\
 32   & -10.0273  &   -0.2612  &    3.1813    &    -7.1072\,(3) & -6.8299 & -22.468 \\
 40   & -10.4743  &   -0.4213  &    3.6250    &    -7.2706\,(4) & -6.8508 & -22.594 \\
 54   & -11.6370  &   -0.8099  &    4.7900    &    -7.6569\,(5) & -6.9540 & -22.856 \\
 70   & -13.8228  &   -1.4897  &    7.0179    &    -8.2945\,(5) & -7.2907 & -23.193 \\
 82   & -16.4083  &   -2.2884  &    9.7062    &    -8.9905\,(6) & -7.8666 & -23.490 \\
 83   & -16.6750  &   -2.3725  &    9.9863    &    -9.0613\,(6) & -7.9348 & -23.519  \\
 92   & -19.5455  &   -3.3111  &   13.0311    &    -9.8254\,(7) & -8.7545 & -23.837 \\
\end{tabular}
\end{ruledtabular}
$^a\,$ Volotka {\em et al.} 2014 \cite{volotka:14}.
\end{table*}

\begin{table*}[t]
\caption{Binding corrections to the $g$-factor of the ground state of Li-like ions, in $10^{-6}$.
The sum of all binding contributions is the difference of the atomic $g$ factor and the
free-electron $g$ factor, $g_e = 2.002\,319\,304\,361\,5\,(6)$
\cite{hanneke:08}.
\label{tab:gtotal}}
\begin{ruledtabular}
\begin{tabular}{llw{4.8}w{4.8}w{4.8}w{4.8}}
\multicolumn{1}{l}{Effect}
& \multicolumn{1}{c}{Contribution}
  & \multicolumn{1}{c}{$^{ 12}$C$^{  3+}$}
                    & \multicolumn{1}{c}{$^{ 16}$O$^{  5+}$}
                                & \multicolumn{1}{c}{$^{ 28}$Si$^{ 11+}$}
                                            &  \multicolumn{1}{c}{$^{ 40}$Ca$^{ 17+}$}
  \\
\hline\\[-5pt]
\multicolumn{1}{l}{Electronic structure}
&$1/Z^0$      & -319.6997      & -568.6205      & -1745.2493        & -3573.9891 \\
&$1/Z^1$      &  137.4194      &  183.3202      &   321.5908        &   461.1500 \\
&$1/Z^2$      &   -6.8363\,(3) &   -6.8436\,(2) &    -6.8787\,(1)   &    -6.9341\,(3) \\
&$1/Z^{3+}$   &    0.1478\,(6) &    0.1377\,(8) &     0.0942\,(15)  &     0.0695\,(25) \\[3pt]
\multicolumn{1}{l}{One-loop QED}
&$1/Z^0$      &  0.1978        &  0.3629        &    1.2244         &  2.7349 \\
&$1/Z^1$      & -0.0974\,(7)   & -0.1329\,(5)   &   -0.2460\,(6)    & -0.3675\,(6) \\
&$1/Z^{2+}$   &  0.0091\,(1)   &  0.0092\,(2)   &    0.0096\,(6)    &  0.0100\,(11) \\[3pt]
\multicolumn{1}{l}{Recoil}
&$1/Z^0$      &  0.0219        &  0.0293       &     0.0515         &  0.0742 \\
&$1/Z^{1}$    & -0.0075        & -0.0076       &    -0.0076         & -0.0076 \\
&$1/Z^{2+}$   & -0.0005        & -0.0004       &    -0.0003         & -0.0002 \\[3pt]
\multicolumn{1}{l}{Two-loop QED}
&$1/Z^0$      & -0.0003        & -0.0005        &   -0.0017         & -0.0044\,(3) \\
&$1/Z^1$      &  0.0001        &  0.0002        &    0.0003\,(1)    &  0.0005\,(4) \\[3pt]
\multicolumn{1}{l}{Finite nuclear size}
&$1/Z^0$      & 0.0001        &  0.0002         &    0.0026         &  0.0144 \\
&$1/Z^1$      &               & -0.0001         &   -0.0005         & -0.0020\,(1) \\[3pt]
%
%
\multicolumn{1}{l}{Radiative recoil}
&$1/Z^0$      &    &    &    & -0.0001 \\[3pt]
%
%
%
%
%
%
\multicolumn{1}{l}{Total theory}
&$g-g_e$      &        -188.8455\,(10) &        -391.7458\,(10) &       -1429.4107\,(17)    &       -3117.2515\,(27) \\[1pt]
&$g$          & 2\,002\,130.4588\,(10) & 2\,001\,927.5585\,(10) & 2\,000\,889.8937\,(17)    & 1\,999\,202.0529\,(27) \\[1pt]
\multicolumn{1}{l}{Previous theory}
&$g$          & 2\,002\,130.457\,(5)^a & 2\,001\,927.558\,(10)^a& 2\,000\,889.8944\,(34)^b  & 1\,999\,202.041\,(13)^c \\[1pt]

\multicolumn{1}{l}{Experiment}
& $g$         &                    &                            & 2\,000\,889.88845\,(14)^b & 1\,999\,202.0406\,(11)^d\\[1pt]
&             &                    &                            & 2\,000\,889.8884\,(19)^e
\end{tabular}
\end{ruledtabular}
$^a$ Yerokhin {\em et al.} 2017 \cite{yerokhin:17:gfact}; $^b$ Glazov {\em et al.} 2019
\cite{glazov:19}; $^c$ Volotka {\em et al.} 2014 \cite{volotka:14};  $^d$ K\"ohler {\em et al.}
2016 \cite{koehler:16};  $^e$ Wagner {\em et al.} 2013 \cite{wagner:13}.
\end{table*}

\section{Results and discussion}

Numerical results of our calculations of the two-photon exchange corrections for the ground state
of Li-like ions are presented in Tables~\ref{tab:2ph:Si} and \ref{tab:2ph}.
Table~\ref{tab:2ph:Si} contains a breakdown of our calculations in two gauges for $Z = 14$ and $Z
= 83$ and demonstrates the gauge invariance of our numerical results. Table~\ref{tab:2ph}
presents the final results of our calculations for $Z = 6$ -- $ 92$. It also compares results of
the QED calculation with those obtained within the standard MBPT and the MBPT with the partial
inclusion of the negative-energy spectrum (``MBPT-neg''). We observe that the standard MBPT
yields the two-photon exchange correction by about three times larger than the complete QED
results. The disagreement is evidently present even in the limit of $Z \to 0$. On the contrary,
the MBPT-neg approach closely reproduces the QED treatment in the region of low values of $Z$.
Previously the same conclusion was reached by the St.~Petersburg group
\cite{volotka:priv,wagner:13}.

For low values of the nuclear charge $Z$, results of our QED and MBPT-neg calculations can be
compared with the prediction of the nonrelativistic QED (NRQED) theory based on the explicitly
correlated three-electron wave function \cite{yerokhin:17:gfact}. According to
Ref.~\cite{yerokhin:17:gfact}, the two-photon exchange correction in the  limit $Z\to 0$ is given
by (see Eq. (16) of that work)
\begin{align}
\Delta g_{\rm 2ph}(Z = 0) = -0.128\,204\,(9)\,\alpha^2 = -6.8270\,(5)\times 10^{-6}\,.
\end{align}
Fig.~\ref{fig:2ph} shows the comparison of the QED, MBPT-neg, and NRQED results. We observe that
all three methods yield results converging to each other in the limit $Z\to 0$.  The difference
between the results by different methods scales as $\propto (\Za)^2$ as expected.  It is
interesting that the MBPT-neg approach does not yield any significant improvement over the NRQED
treatment for low- and medium-$Z$ ions.

The QED calculation of the two-photon exchange correction for the $g$ factor of Li-like ions was
previously carried out in Refs.~\cite{volotka:12,volotka:14}. Unfortunately, the numerical
results were presented only for four ions and mostly in the form of the total electron-electron
interaction correction. The only ion for which the calculations are directly comparable is
silicon, $Z = 14$, for which we find some tension. Our calculation yields $-6.8787\,(1) \times
10^{-6}$, whereas Ref.~\cite{volotka:14} reported $-6.876 \times 10^{-6}$. As an additional
cross-check, we performed calculations for the two-photon exchange correction to the ground-state
hyperfine splitting of Li-like bismuth (which is another example of a magnetic perturbation
potential) and found agreement with results listed in Table~I of Ref.~\cite{volotka:12}.

Having obtained results for the two-photon exchange correction, we are now in a position to
update the theoretical predictions for the ground-state $g$ factor of Li-like ions.
Table~\ref{tab:gtotal} presents a compilation of all known binding corrections to the $g$ factor
of the ground state of four Li-like ions, C$^{3+}$, O$^{5+}$, Si$^{11+}$, and Ca$^{ 17+}$. As
compared to the analogous compilation in our previous investigation \cite{yerokhin:17:gfact}, we
introduced several improvements. The two-photon exchange correction (i.e., the
electronic-structure contribution of relative order $1/Z^2$) is computed in the present work.
Beside this, we included the one-loop $1/Z$ and $1/Z^{2+}$ QED effects from our recent work
\cite{yerokhin:20:gfact} and the nuclear recoil corrections of relative orders $1/Z^0$, $1/Z^1$,
and $1/Z^{2+}$ calculated by Shabaev {\em et al.}~\cite{shabaev:17:prl}. Furthermore, we added
the two-loop $(\Za)^5$ effects calculated recently by Czarnecki and co-workers
\cite{czarnecki:18,czarnecki:20} and by us \cite{yerokhin:13:twoloopg}. The two-loop results in
those studies were reported for the $1s$ hydrogenic state. We here convert them to the $2s$ state
by assuming the $1/n^3$ scaling. Having in mind that the result for the nonlogarithmic $(\Za)^5$
term is not complete, we ascribe the uncertainty of 20\% to the two-loop $(\Za)^5$ correction.
The uncertainty due to higher-order two-loop effects was evaluated on the basis of available
one-loop results, with the extension factor of 2.

One of the largest uncertainties of the theoretical predictions comes from the higher-order
electronic-structure correction $\sim\!1/Z^{3+}$. The values in Table~\ref{tab:gtotal} for this
correction are obtained within NRQED in Ref.~\cite{yerokhin:17:gfact}. Theoretical estimates for
their uncertainties are obtained by taking the relative deviation of the NRQED and full QED
results for the $1/Z^2$ correction and multiplying it by the extension factor of 2.

The comparison presented in Table~\ref{tab:gtotal} shows agreement of the present theoretical
$g$-factor values with previous theoretical predictions. In particular, for C$^{3+}$ and
O$^{5+}$, our results are in excellent agreement with, but 5-10 times more precise than our
earlier results in Ref.~\cite{yerokhin:17:gfact}. For Si$^{11+}$ and Ca$^{17+}$, our total
$g$-factor values are in agreement with the previous theoretical results of the St.~Petersburg
group \cite{volotka:14,glazov:19}. We note, however, that some tension exists between the
calculations on the level of individual contributions. Specifically, the total electron-electron
interaction correction for silicon in Ref.~\cite{glazov:19} is reported as $314.812\,(3)\times
10^{-6}$, whereas our calculation yields $314.806\,(2)\times 10^{-6}$. This deviation disappears
when the electron-structure correction is combined with the $1/Z$ QED contribution.

Table~\ref{tab:gtotal} also compares the obtained theoretical predictions with experimental
results available for two Li-like ions, Si$^{11+}$ and Ca$^{ 17+}$. In both cases theoretical
values deviate from the experimentally observed $g$ factors, by 3.1$\,\sigma$ for silicon and
4.2$\,\sigma$ for calcium. The discrepancy grows with the increase of $Z$. Such effect could be
caused by some unknown contribution missing in theoretical calculations.

The largest uncertainty in the theoretical predictions for Li-like silicon and calcium presently
comes from the higher-order electron correction $\sim\!1/Z^{3}$. This contribution cannot be
calculated rigorously to all orders in $\Za$ but needs to be treated by approximate methods or
within the $\Za$ expansion. It should be pointed out that our calculations and those by the
St.~Petersburg group \cite{volotka:14,glazov:19} use different approaches for handling this
correction. Our result is based on the $\Za$ expansion, whereas the St.~Petersburg group used
screening potentials in the two-photon-exchange calculations and explicitly computed the
three-photon-exchange contribution within the Breit approximation \cite{glazov:19}. The
$\Za$-expansion approach is most suitable in the low-$Z$ region, whereas the screening-potential
method is advantageous for high-$Z$ ions. For light ions, the $\Za$-expansion results can be
improved further,  by performing the NRQED calculation of the next-order term in the $\alpha$
expansion.

Summarizing, we performed calculations of the two-photon exchange corrections to the $g$ factor
of the ground state of Li-like ions without an expansion in the nuclear binding strength
parameter $Z\alpha$. The calculations were carried out in two gauges, the Feynman and the Coulomb
ones, thus allowing an explicit test of the gauge invariance. In the low-$Z$ region, the obtained
results were checked against those delivered by two different and independent methods, namely,
the relativistic many-body perturbation theory with a partial inclusion of the negative-energy
continuum and the nonrelativistic quantum electrodynamics. It was demonstrated that all three
methods yield consistent results in the limit of small nuclear charges.

Our calculation improves the overall accuracy of theoretical predictions of the $g$ factor of
Li-like ions, especially in the low-$Z$ region. An agreement with previous theoretical
calculations is found. However, the theoretical predictions are shown to systematically deviate
from the experimental results for Li-like silicon and calcium, by approx.~3 and 4 standard
deviations, respectively. The reason for these discrepancies is not known at present, but is
likely to be on the theoretical side. We conclude that further work is needed in order to find
the reasons behind the observed discrepancies.

\begin{acknowledgments}
Work of V.A.Y. is supported by the Russian Science Foundation (Grant No. 20-62-46006). Z.H. and
C.H.K. are supported by the Deutsche Forschungsgemeinschaft (DFG, German Research Foundation) –
Project-ID 273811115 – SFB 1225.
\end{acknowledgments}

\appendix
\section{Pole terms and subtractions in principal-value integrals}
\label{sec:subtractions}

In this section we present explicit formulas used in the present work to numerically evaluate
integrals with poles separated by the infinitesimal small additions from the integration contour.
The evaluation procedure is as follows. First, use the Sokhotsky-Plemelj formula
\begin{align}
\frac1{z+i0} = {\cal P}\frac1{z} - i\pi \delta(z)\,,
\end{align}
to convert integrals with poles near the integration contour to the principal-value integrals.
Next, we expand the integrand in a Taylor series in the vicinity of the poles and determine the
subtractions that remove singularities from the integrand. Finally, we re-add the subtractions
and the perform the principal-value integrals analytically. In the Lamb-shift calculations, we
encounter the integrals of three types, evaluated as follows
\begin{widetext}
\begin{align}\label{eq:A1}
\int_a^b d\omega\, \frac{F(\omega)}{(\Delta_1 -\omega+i0)(\Delta_2+\omega+i0)}
     & \ \underset{\scriptscriptstyle \Delta_1\ne -\Delta_2}{=}
 -i\pi \, \frac{F(\Delta_1)}{\Delta_1 + \Delta_2}\,\delta_{\Delta_1 \in (ab)}
 -i\pi \, \frac{F(-\Delta_2)}{\Delta_1 + \Delta_2}\,\delta_{-\Delta_2 \in (ab)}
 \nonumber \\
  &\
 + \frac{F(\Delta_1)}{\Delta_1 + \Delta_2}\,\ln \left| \frac{a-\Delta_1}{b-\Delta_1}\right|
 + \frac{F(-\Delta_2)}{\Delta_1 + \Delta_2}\,\ln \left| \frac{b+\Delta_2}{a+\Delta_2}\right|
 \nonumber \\
  &\
+ \int_a^b d\omega\, \bigg[
 \frac{F(\omega)}{(\Delta_1 -\omega)(\Delta_2+\omega)}
- \frac{F(\Delta_1)}{(\Delta_1 +\Delta_2)(\Delta_1-\omega)}
- \frac{F(-\Delta_2)}{(\Delta_1 +\Delta_2)(\Delta_2+\omega)}
\bigg]\,,
\end{align}

\begin{align}
\int_a^b d\omega\, \frac{F(\omega)}{(\Delta_1 -\omega+i0)(\Delta_2-\omega+i0)}
   &\  \underset{\scriptscriptstyle \Delta_1\ne \Delta_2}{=}
 -i\pi \, \frac{F(\Delta_1)}{\Delta_2-\Delta_1}\,\delta_{\Delta_1 \in (ab)}
 -i\pi \, \frac{F(\Delta_2)}{\Delta_1-\Delta_2}\,\delta_{\Delta_2 \in (ab)}
 \nonumber \\
  &\
 + \frac{F(\Delta_1)}{\Delta_2 - \Delta_1}\,\ln \left| \frac{a-\Delta_1}{b-\Delta_1}\right|
 + \frac{F(\Delta_2)}{\Delta_1 - \Delta_2}\,\ln \left| \frac{a-\Delta_2}{b-\Delta_2}\right|
 \nonumber \\
  &\
+ \int_a^b d\omega\, \bigg[
 \frac{F(\omega)}{(\Delta_1 -\omega)(\Delta_2-\omega)}
- \frac{F(\Delta_1)}{(\Delta_2-\Delta_1)(\Delta_1-\omega)}
- \frac{F(\Delta_2)}{(\Delta_1-\Delta_2)(\Delta_2-\omega)}
\bigg]\,,
\end{align}
\begin{align}
\int_a^b d\omega\, \frac{F(\omega)}{(\Delta -\omega+i0)^2}
 = &\
 i\pi \, F'(\Delta)\,\delta_{\Delta \in (ab)}
 + F(\Delta)\Big[\frac{1}{a-\Delta}-\frac{1}{b-\Delta}\Big]
 + F'(\Delta)\,\ln \left| \frac{b-\Delta}{a-\Delta}\right|
 \nonumber \\
  &\
+ \int_a^b d\omega\,
 \frac{F(\omega)-F(\Delta)-(\omega-\Delta)\, F'(\Delta)}{(\Delta -\omega)^2}
\,.
\end{align}
\end{widetext}
Here, $\delta_{\Delta \in (ab)}$ is 1 if $\Delta \in (ab)$ and zero otherwise. Note that the
integrals in the right-hand-side of the above identities are regular, without any need to assume
the principal value. In order to determine the subtractions in the integrand, we use, e.g., for
Eq.~(\ref{eq:A1}), that
\begin{align}
\frac{F(\omega)}{(\Delta_1 -\omega)(\Delta_2+\omega)} = \frac1{\Delta_1+\Delta_2}\,
\bigg( \frac{F(\omega)}{\Delta_1 -\omega} + \frac{F(\omega)}{\Delta_2 +\omega}\bigg)\,,
\end{align}
and then expanded the terms in the right-hand side in the vicinity of the corresponding poles.

Calculations for the $g$ factor require formulas with higher powers of denominators. Such
formulas can be obtained by formal differentiation of the above identities over $\Delta$'s.

\section{Matrix elements of the electron-electron interaction in Coulomb gauge}
\label{app:coul}

The electron-electron interaction operator in the Coulomb gauge is given by
Eq.~(\ref{eq:I:coul}). The matrix element of this operator is conveniently expressed in the
standard two-body-operator form that separates the angular and radial parts,
\begin{align}
\lbr ab | I_{\rm Coul}(\omega)| cd \rbr = \sum_L J_L(abcd)\, {R}^{\rm Coul}_{L}(\omega,abcd)\,,
\end{align}
where $J_L$ contains the dependence on the angular-momentum projections,
\begin{align}
  J_L(abcd) =&\  \sum_{m_L} \frac{(-1)^{L-m_L+j_c-m_c+j_d-m_d}}{2L+1}\,
  \nonumber \\ & \times
  C^{Lm_L}_{j_am_a,j_c-m_c} C^{Lm_L}_{j_dm_d,j_b-m_b}\,,
\end{align}
with $C_{l_1m_1,l_2m_2}^{lm}$ being the Clebsch-Gordan coefficients, and $R_L^{\rm Coul}$ is the
radial integral. Formulas for the radial integral in the Coulomb gauge were derived in
Ref.~\cite{mann:71}. For our purposes it is convenient to express them in a form similar to the
Feynman-gauge radial integrals \cite{yerokhin:20:green},
\begin{widetext}
\begin{align}\label{eq:Rcoul}
R^{\rm Coul}_{J}(\omega,abcd) = &\ \int_0^{\infty}dx_1dx_2\, (x_1x_2)^2\,
   \bigg\{
     (2J+1)(-1)^J\,C_J(\kappa_a,\kappa_c)\,C_J(\kappa_b,\kappa_d)\,
       g_J(0,x_1,x_2)\, W_{ac}(x_1)\, W_{bd}(x_2)
 \nonumber \\
  &
 - \sum_{L = J-1}^{J+1} (-1)^L\, a_{JL}\, g_L(\omega,x_1,x_2)\, X_{ac,JL}(x_1)\, X_{bd,JL}(x_2)
 \nonumber \\
  &
- (-1)^J\sqrt{J(J+1)}\, \Big[
  g^{\rm ret}_J(\omega,x_1,x_2)\, X_{ac,JJ+1}(x_1)\, X_{bd,JJ-1}(x_2)
\nonumber \\ &
+  g^{\rm ret}_J(\omega,x_2,x_1)\, X_{ac,JJ-1}(x_1)\, X_{bd,JJ+1}(x_2)\Big] \bigg\}\,,
\end{align}
where $g_l(\omega,x_1,x_2) = i\omega\,j_l(\omega x_<)\,h^{(1)}_l(\omega x_>)$, the coefficients
$a_{JL}$ are given by
\begin{align}
a_{JL} = \left\{
        \begin{array}{cl}
        J+1 \,,  & {\mbox{\rm when} \ \ } L = J-1 \,,\\
        2J+1 \,, & {\mbox{\rm when} \ \ } L = J \,,\\
        J   \,,  & {\mbox{\rm when} \ \ } L = J+1\,,
        \end{array}
        \right.
\end{align}
and
\begin{align}
g^{\rm ret}_l(\omega,x_1,x_2)= \left\{
        \begin{array}{cl}
         i\omega\,j_{l+1}(\omega x_<)\,h^{(1)}_{l-1}(\omega x_>) \,,  & {\mbox{\rm when} \ \ } x_1 < x_2 \,,\\[5pt]
         i\omega\,j_{l-1}(\omega x_<)\,h^{(1)}_{l+1}(\omega x_>)
            - \frac{\displaystyle 2l+1}{\displaystyle \omega^2}\frac{\displaystyle x_<^{l-1}}{\displaystyle x_>^{l+2}} \,,  & {\mbox{\rm when} \ \ } x_1 > x_2 \,.\\
        \end{array}
        \right.
\end{align}
Furthermore,
\begin{align}
X_{ac,ll'}(r) = & \ g_a(r)\,f_c(r)\,S_{ll'}(-\kappa_c,\kappa_a) + f_a(r)\,g_c(r)\,S_{ll'}(\kappa_c,-\kappa_a)
 \,, \\
W_{ac}(r) = & \ g_a(r)\,g_c(r) + f_a(r)\,f_c(r)
 \,,
\end{align}
and the standard angular coefficients $S_{ll'}$ and $C_l$ are defined by Eqs.~(A7)-(A10) of
Ref.~\cite{yerokhin:20:green}.

We note that the function $g^{\rm ret}_l$ has a finite limit at $\omega\to 0$, as the
$1/\omega^2$ term cancels with the first term of the small-argument expansion of the spherical
Bessel functions. The limiting form is
\begin{align}
g^{\rm ret}_l(0,x_1,x_2)= \left\{
        \begin{array}{cl}
         0 \,,  & {\mbox{\rm when} \ \ } x_1 < x_2 \,,\\[5pt]
  -\nicefrac12
            \left ( \frac{\displaystyle x_<^{l-1}}{\displaystyle x_>^{l}}
            - \frac{\displaystyle x_<^{l+1}}{\displaystyle x_>^{l+2}}
            \right)
             \,,  & {\mbox{\rm when} \ \ } x_1 > x_2 \,.\\
        \end{array}
        \right.
\end{align}
The presence of the spurious singularity at $\omega \to 0$ leads to numerical instabilities in
the computation $g^{\rm ret}_l$ at small $\omega$, especially when evaluating the derivatives
$I'(\omega)$ and $I''(\omega)$. In order to facilitate computations for small $\omega$, we
introduce regularized functions $\overline{j}_l$ and $\overline{h}_l$, separating the first term
of the small-argument expansion, as follows
\begin{align}
j_l(z) \equiv &\ \frac{z^l}{(2l+1)!!} +\overline{j}_l(z)\,,\\
h^{(1)}_l(z) \equiv &\ \frac{(2l-1)!!}{i\,z^{l+1}} +\overline{h}_l^{(1)}(z)\,.
\end{align}
We thus obtain a regular representation for $g^{\rm ret}_l$, which is suitable for a numerical
evaluation,
\begin{align}
g^{\rm ret}_l(\omega,x_1,x_2)= i\omega\, \left\{
        \begin{array}{cl}
         j_{l+1}(\omega x_<)\,h^{(1)}_{l-1}(\omega x_>) \,,  & {\mbox{\rm when} \ \ } x_1 < x_2 \,,\\[5pt]
         \overline{j}_{l-1}(\omega x_<)\,h^{(1)}_{l+1}(\omega x_>)
        +   \frac{\displaystyle (\omega x_<)^{l-1}}{\displaystyle (2l-1)!!} \,\overline{h}^{(1)}_{l+1}(\omega x_>)
         \,,  & {\mbox{\rm when} \ \ } x_1 > x_2 \,.\\
        \end{array}
        \right.
\end{align}
In the computation of the second derivative of $g^{\rm ret}_l$ over $\omega$, we had to separate
out two first terms of the the small-argument expansion of the spherical Bessel functions, in
order to achieve an explicit cancelation of singular terms.

We note that when the matrix element $\lbr ab| I_{\rm Coul}(\omega)|cd\rbr$ is calculated with
eigenfunctions of the one-particle Dirac Hamiltonian $h_D$, it can be simplified by using the
commutator relation $ -i\,\balpha\cdot \bnabla\, e^{i\omega x_{12}} = \big[ h_D, e^{i\omega
x_{12}}\big] $, where  $[\,,]$ denotes commutator. In this case, we immediately have
\begin{align}\label{eq:B13}
\lbr ab|  I_{\rm Coul}(\omega)|cd\rbr = \alpha\,
   \lbr ab| \left[ \frac1{x_{12}}
  -  \balpha_1\cdot\balpha_2\,
   \frac{e^{\im|\omega|x_{12}}}{x_{12}}
   - \frac{(\vare_a-\vare_c)(\vare_b-\vare_d)}{\omega^2}
    \frac{e^{\im|\omega|x_{12}}-1}{x_{12}}
    \right]|cd\rbr \,,
\end{align}
and thus the Coulomb-gauge matrix element is expressed in terms of the Feynman-gauge matrix
elements. The above expression is convenient by its simplicity but it has a spurious singularity
at $\omega \to 0$ that might lead to numerical instabilities in practical calculations. This form
of the Coulomb matrix element proved to be very useful for demonstrating the gauge invariance of
photon-exchange corrections \cite{soguel:21}.

\end{widetext}


\begin{thebibliography}{10}

\bibitem{sturm:11} S.~Sturm, A.~Wagner, B.~Schabinger, J.~Zatorski, Z.~Harman, W.~Quint,
    G.~Werth,
  C.~H. Keitel, and K.~Blaum,
\newblock Phys. Rev. Lett. {\bf 107}, 023002 (2011).

\bibitem{koehler:16} F.~K{\"o}hler, K.~Blaum, M.~Block, S.~Chenmarev, S.~Eliseev, D.~A. Glazov,
  M.~Goncharov, J.~Hou, A.~Kracke, D.~A. Nesterenko, Y.~N. Novikov, W.~Quint,
  E.~Minaya~Ramirez, V.~M. Shabaev, S.~Sturm, A.~V. Volotka, and G.~Werth,
\newblock Nat. Comm. {\bf 7}, 10246 (2016).

\bibitem{arapoglou:19} I.~Arapoglou, A.~Egl, M.~H\"ocker, T.~Sailer, B.~Tu, A.~Weigel, R.~Wolf,
  H.~Cakir, V.~A. Yerokhin, N.~S. Oreshkina, V.~A. Agababaev, A.~V. Volotka,
  D.~V. Zinenko, D.~A. Glazov, Z.~Harman, C.~H. Keitel, S.~Sturm, and K.~Blaum,
\newblock Phys. Rev. Lett. {\bf 122}, 253001 (2019).

\bibitem{sturm:14} S.~Sturm, F.~K\"ohler, J.~Zatorski, A.~Wagner, Z.~Harman, G.~Werth, W.~Quint,
  C.~H. Keitel, and K.~Blaum,
\newblock Nature {\bf 506}, 467–470 (2014).

\bibitem{glazov:19} D.~A. Glazov, F.~K\"ohler-Langes, A.~V. Volotka, K.~Blaum, F.~Hei\ss{}e,
  G.~Plunien, W.~Quint, S.~Rau, V.~M. Shabaev, S.~Sturm, and G.~Werth,
\newblock Phys. Rev. Lett. {\bf 123}, 173001 (2019).

\bibitem{mohr:16:codata} P.~J. Mohr, D.~B. Newell, and B.~N. Taylor,
\newblock Rev. Mod. Phys. {\bf 88}, 035009 (2016).

\bibitem{sturm:13:Si} S.~Sturm, A.~Wagner, M.~Kretzschmar, W.~Quint, G.~Werth, and K.~Blaum,
\newblock Phys. Rev. A {\bf 87}, 030501 (2013).

\bibitem{vogel:19} M.~Vogel, M.~S. Ebrahimi, Z.~Guo, A.~Khodaparast, G.~Birkl, and W.~Quint,
\newblock Annalen der Physik {\bf 531}, 1800211 (2019).

\bibitem{shabaev:06:prl} V.~M. Shabaev, D.~A. Glazov, N.~S. Oreshkina, A.~V. Volotka, G.~Plunien,
    H.-J.
  Kluge, and W.~Quint,
\newblock Phys. Rev. Lett. {\bf 96}, 253002 (2006).

\bibitem{yerokhin:16:gfact:prl} V.~A. Yerokhin, E.~Berseneva, Z.~Harman, I.~I. Tupitsyn, and
    C.~H. Keitel,
\newblock Phys. Rev. Lett. {\bf 116}, 100801 (2016).

\bibitem{debierre:20} V.~Debierre, C.~Keitel, and Z.~Harman,
\newblock Phys. Lett. B {\bf 807}, 135527 (2020).

\bibitem{shabaev:02:li} V.~M. Shabaev, D.~A. Glazov, M.~B. Shabaeva, V.~A. Yerokhin, G.~Plunien,
    and
  G.~Soff,
\newblock Phys. Rev. A {\bf 65}, 062104 (2002).

\bibitem{yerokhin:16:gfact:pra} V.~A. Yerokhin, E.~Berseneva, Z.~Harman, I.~I. Tupitsyn, and
    C.~H. Keitel,
\newblock Phys. Rev. A {\bf 94}, 022502 (2016).

\bibitem{cakir:20} H.~Cakir, V.~A. Yerokhin, N.~S. Oreshkina, B.~Sikora, I.~I. Tupitsyn, C.~H.
  Keitel, and Z.~Harman,
\newblock Phys. Rev. A {\bf 101}, 062513 (2020).

\bibitem{volotka:12} A.~V. Volotka, D.~A. Glazov, O.~V. Andreev, V.~M. Shabaev, I.~I. Tupitsyn,
    and
  G.~Plunien,
\newblock Phys. Rev. Lett. {\bf 108}, 073001 (2012).

\bibitem{volotka:14} A.~V. Volotka, D.~A. Glazov, V.~M. Shabaev, I.~I. Tupitsyn, and G.~Plunien,
\newblock Phys. Rev. Lett. {\bf 112}, 253004 (2014).

\bibitem{yerokhin:17:gfact} V.~A. Yerokhin, K.~Pachucki, M.~Puchalski, Z.~Harman, and C.~H.
    Keitel,
\newblock Phys. Rev. A {\bf 95}, 062511 (2017).

\bibitem{yerokhin:10:sehfs} V.~A. Yerokhin and U.~D. Jentschura,
\newblock Phys. Rev. A {\bf 81}, 012502 (2010).

\bibitem{shabaev:94:ttg1} V.~M. Shabaev and I.~G. Fokeeva,
\newblock Phys. Rev. A {\bf 49}, 4489  (1994).

\bibitem{yerokhin:01:2ph} V.~A. Yerokhin, A.~N. Artemyev, V.~M. Shabaev, M.~M. Sysak, O.~M.
    Zherebtsov,
  and G.~Soff,
\newblock Phys. Rev. A {\bf 64}, 032109 (2001).

\bibitem{volotka:priv} A.~V. Volotka,
\newblock private communication, 2014.

\bibitem{wagner:13} A.~Wagner, S.~Sturm, F.~K\"ohler, D.~A. Glazov, A.~V. Volotka, G.~Plunien,
  W.~Quint, G.~Werth, V.~M. Shabaev, and K.~Blaum,
\newblock Phys. Rev. Lett. {\bf 110}, 033003 (2013).

\bibitem{mohr:00:pra} P.~J. Mohr and J.~Sapirstein,
\newblock Phys. Rev. A {\bf 62}, 052501 (2000).

\bibitem{kosheleva:20} V.~P.~Kosheleva, A.~V.~Volotka, D.~A.~Glazov,  and S.~Fritzsche,
\newblock Phys. Rev. Research {\bf 2}, 013364 (2020).

\bibitem{shabaev:04:DKB} V.~M. Shabaev, I.~I. Tupitsyn, V.~A. Yerokhin, G.~Plunien, and G.~Soff,
\newblock Phys. Rev. Lett. {\bf 93}, 130405 (2004).

\bibitem{angeli:13} I.~Angeli and K.~Marinova,
\newblock At. Dat. Nucl. Dat. Tabl. {\bf 99}, 69  (2013).

\bibitem{hanneke:08} D.~Hanneke, S.~Fogwell, and G.~Gabrielse,
\newblock Phys. Rev. Lett. {\bf 100}, 120801 (2008).

\bibitem{yerokhin:20:gfact} V.~A. Yerokhin, K.~Pachucki, M.~Puchalski, C.~H. Keitel, and
    Z.~Harman,
\newblock Phys. Rev. A {\bf 102}, 022815 (2020).

\bibitem{shabaev:17:prl} V.~M. Shabaev, D.~A. Glazov, A.~V. Malyshev, and I.~I. Tupitsyn,
\newblock Phys. Rev. Lett. {\bf 119}, 263001 (2017).

\bibitem{czarnecki:18} A.~Czarnecki, M.~Dowling, J.~Piclum, and R.~Szafron,
\newblock Phys. Rev. Lett. {\bf 120}, 043203 (2018).

\bibitem{czarnecki:20} A.~Czarnecki, J.~Piclum, and R.~Szafron,
\newblock Phys. Rev. A {\bf 102}, 050801 (2020).

\bibitem{yerokhin:13:twoloopg} V.~A. Yerokhin and Z.~Harman,
\newblock Phys. Rev. A {\bf 88}, 042502 (2013).

\bibitem{mann:71} J.~B. Mann and W.~R. Johnson,
\newblock Phys. Rev. A {\bf 4}, 41 (1971).

\bibitem{yerokhin:20:green} V.~A. Yerokhin and A.~V. Maiorova,
\newblock Symmetry {\bf 12}, 800 (2020).

\bibitem{soguel:21} R.~N. Soguel, A.~V. Volotka, E.~V. Tryapitsyna, D.~A. Glazov, V.~P.
    Kosheleva,
  and S.~Fritzsche,
\newblock Phys. Rev. A {\bf 103}, 042818 (2021).

\end{thebibliography}

\end{document}